# Orbital-Exchange and Fractional Quantum Number Excitations in an f-electron Metal Yb$_2$Pt$_2$Pb

L. S. Wu,[1,2,3] W. J. Gannon,[1,2,4] I. A. Zaliznyak,[2,*] A. M. Tsvelik,[2] M. Brockmann,[5,6] J.-S. Caux,[5] M. S. Kim,[2] Y. Qiu,[7] J. R. D. Copley,[7] G. Ehlers,[3] A. Podlesnyak,[3] and M. C. Aronson[1,2,4]

[1]Department of Physics and Astronomy, Stony Brook University, Stony Brook, NY 11794 USA

[2]Condensed Matter Physics and Materials Science Division,
Brookhaven National Laboratory, Upton, NY 11973 USA

[3]Quantum Condensed Matter Division, Oak Ridge National Laboratory, Oak Ridge 37831 TN USA

[4]Department of Physics and Astronomy, Texas A& M University, College Station, TX 77843, USA

[5]Institute for Theoretical Physics, University of Amsterdam,
Science Park 904, 1098 XH Amsterdam, The Netherlands

[6]Max Planck Institute for the Physics of Complex Systems,
Nöthnitzer Str. 38, 01187 Dresden, Germany

[7]NIST Center for Neutron Research, National Institute of
Standards and Technology, Gaithersburg, MD 20899 USA

(Dated: March 12, 2016)

_______
* Corresponding author: zaliznyak@bnl.gov

**Exotic quantum states and fractionalized magnetic excitations, such as spinons in one-dimensional chains, are generally viewed as belonging to the domain of 3d transition metal systems with spins 1/2. Our neutron scattering experiments on the 4f-electron metal $Yb_2Pt_2Pb$ overturn this common wisdom. We observe broad magnetic continuum dispersing in only one direction, which indicates that the underlying elementary excitations are spinons carrying fractional spin-1/2. These spinons are the quantum dynamics of the anisotropic, orbital-dominated Yb moments, and thus these effective quantum spins are emergent variables that encode the electronic orbitals. The unique birthmark of their unusual origin is that only longitudinal spin fluctuations are measurable, while the transverse excitations such as spin waves are virtually invisible to magnetic neutron scattering. The proliferation of these orbital-spinons strips the electrons of their orbital identity, and we thus report here a new electron fractionalization phenomenon, charge-orbital separation.**

It is generally believed that fractional quantum excitations such as spinons in one-dimensional (1D) spin chains only proliferate and govern magnetism in systems with small and isotropic atomic magnetic moments, such as spin$-1/2$ $Cu^{2+}$. In contrast, large and anisotropic orbital-dominated moments, such as those produced by strong spin-orbit coupling in the rare earths, are considered to be classical, becoming static as $T \to 0$ since the conventional Heisenberg-Dirac exchange interaction [1, 2] cannot reverse their directions. We present here the results of neutron scattering measurements on $Yb_2Pt_2Pb$ that completely negate this common wisdom. A diffuse continuum of magnetic excitations is observed in $Yb_2Pt_2Pb$, direct evidence that the elementary excitations carry a fractional spin quantum number, $S = 1/2$. The excitations disperse in only one direction, showing that the Yb moments form spin chains that are embedded in, but effectively decoupled from the three-dimensional conduction electron bands in metallic $Yb_2Pt_2Pb$. The spectrum of magnetic excitations strongly resembles the spinon continuum found in $S = 1/2$ Heisenberg spin chains, and indeed comparison to the 1D XXZ Hamiltonian indicates only a moderate exchange anisotropy, $\Delta = J_{zz}/J_{xx} \sim 3$. Here we show how the orbital physics of $4f$-electron exchange interactions can reconcile this moderately-anisotropic quantum Hamiltonian with the extreme anisotropy of the putatively classical Yb ($J = 7/2$) magnetic moments with respect to magnetic fields. We find that the unexpected quantum behavior emerges at low energies from the competition of interactions that act on much higher energy scales, i.e. the strong on-site Coulomb and spin-orbit interactions, as well as the crystal fields, and the inter-site hopping. Our findings thus provide a unique and a hitherto unforeseen manifestation of *emergence* [3] of quantum physics in the system of semi-classical electronic orbitals.

The unusual properties of $Yb_2Pt_2Pb$ derive in part from its crystal structure (Fig. 1A,B), where the $Yb^{3+}$ ions form ladders along the $c-$axis, with rungs on the orthogonal bonds of the Shastry-Sutherland Lattice (SSL) [5] in the $ab$-planes. Equally important is the strong spin-orbit coupling, which combines spin and orbital degrees of freedom into a large, $J = 7/2$ Yb moment. The absence of a Kondo effect indicates minimal coupling of Yb to the conduction electrons of this excellent metal [6, 7]. A point-charge model (Supplementary Information) indicates that the crystal electric field (CEF) lifts the eightfold degeneracy of the $Yb^{3+}$ moments, producing a Kramers doublet ground state that is a nearly pure state of the total angular momentum, $\boldsymbol{J}$, $|J, m_J\rangle = |7/2, \pm 7/2\rangle$. The estimated anisotropy of the Landé $g$-factor is in good agreement with that of the measured magnetization, $g_\parallel/g_\perp = 7.5(4)$ [6–8], implying strong Ising anisotropy in $Yb_2Pt_2Pb$, which confines the individual Yb moments to two orthogonal sublattices in the $ab-$plane. The quantum states of the $|\pm 7/2\rangle$ Ising doublet are the superpositions of its "up" and "down" components, $\alpha_\uparrow|7/2\rangle + \alpha_\downarrow|-7/2\rangle$, and therefore the doublet can be viewed as an effective quantum spin-1/2. However, familiar interactions like the Zeeman, Heisenberg-Dirac exchange, and dipole interactions that are bilinear in $\boldsymbol{J}$ can only change the total angular momentum quantum number by $\Delta m_J = \pm 1$, and so they have no matrix elements that would allow transitions between the moment-reversed states of the ground state wave function. Only multiple virtual processes involving excited states could reverse individual Yb moments, but these processes are very weak since specific heat [6] and inelastic neutron scattering measurements (Supplementary Information) find that the ground and first excited states are separated by as much as 25 meV. Consequently, it would seem that Yb$_2$Pt$_2$Pb would only display static, classical Ising behavior. The reality is, in fact, completely different.

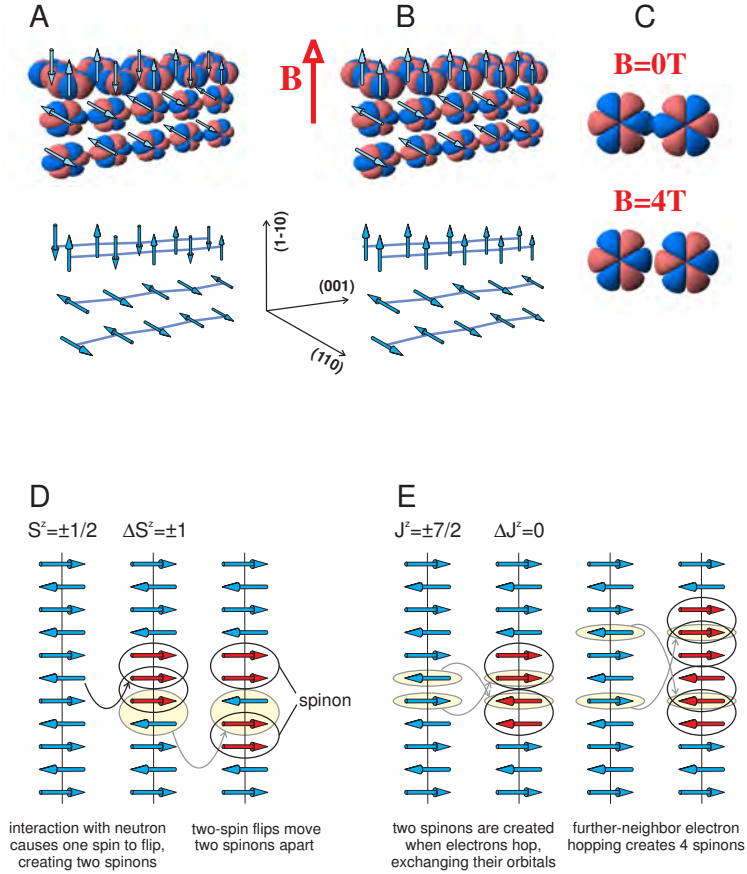

FIG. 1. **Quantum Orbital-Spin Chains in Yb₂Pt₂Pb.** The double chain magnetic structure of Yb$_2$Pt$_2$Pb for $T < T_N = 2.07$ K (A) without magnetic field, and (B) in a 4 T field applied along the (1-10) direction. Arrows indicate the ordered magnetic moment directions, which are parallel to the local Ising easy axes, horizontal for the (110) and vertical for the (1-10) sublattice. Crystal axes are shown by black arrows. Thick red arrow shows magnetic field direction. $|m_L| = 3$ 4f orbitals are shown at 1 ppm electronic density for an effective Slater nuclear charge of $^{70}$Yb [4]. (C) Orbital overlaps for antiferromagnetic (B = 0) and fully saturated (B = 4T) state. (D) Schematic illustration of the two-spinon excitation process via spin flip (magnon) creation in $S = 1/2$ antiferromagnetic chain. Such processes correspond to the change of angular momentum, $\Delta m_S \equiv \Delta S^z = \pm 1$, and are allowed by selection rules that govern interaction with a physical field, such as magnetic field of a neutron, or a photon. (E) For $m_J \equiv J^z = \pm 7/2$ angular momenta in Yb$_2$Pt$_2$Pb flipping magnetic moment requires $\Delta m_J = \pm 7$, and therefore cannot be induced via single-particle processes. The only processes allowed by the selection rules are those with $\Delta m_J = 0$, such as when two electrons hop exchanging their orbitals, i.e. pairwise permutations of electrons. Permutation of two nearest-neighbor electrons creates two spinons, while further-neighbor hopping, such as the permutation of the next-nearest-neighbor electrons in opposite-polarity orbitals illustrated in the figure, results in a four-spinon state.

The neutron scattering experiments on Yb$_2$Pt$_2$Pb that we report here reveal a continuum of low energy quantum excitations that display the distinctive spinon dispersion along the $c-$axis (Fig. 2A), typical of the $S = 1/2$ Heisenberg-Ising XXZ spin Hamiltonian solved by Bethe [9],

$$H = \mathrm{J}\sum_n \left( S_n^x S_{n+1}^x + S_n^y S_{n+1}^y + \Delta S_n^z S_{n+1}^z \right), \tag{1}$$

where J is the Heisenberg spin-exchange coupling, and $\Delta$ is its anisotropy. This observation provides definitive evidence that the Yb moments in Yb$_2$Pt$_2$Pb behave as quantum mechanical spins-1/2 [10]. The spinon spectrum $\mathrm{M}(\mathbf{Q}, E)$ is fully gapped, although the gap is much smaller than the excitation bandwidth, indicating only moderate Ising anisotropy, $\Delta \gtrsim 1$. The lack of any wave vector $Q_{HH}$ dispersion for this gap, as well as for the scattering intensity itself in the $ab-$plane, Fig. 2(B,C), indicates that the dispersing excitations are confined to the ladder rails, which form an array of weakly coupled spin-1/2 chains.

Our next important observation is made by considering the overall wave vector dependence of the energy-integrated intensity $\mathrm{M}(\mathbf{Q})$, Fig. 2(C,D), which reveals that the excitations in each of the two orthogonal sublattices of Yb moments in Yb$_2$Pt$_2$Pb are longitudinally polarized. This is clearly demonstrated in Fig. 2D, where the $\mathrm{M}(\mathbf{Q})$ dependence on $H$ in the $(H, H, L)$ scattering plane is very accurately described by the projections of Yb moments on the wave vector, consistent with the polarization factor in the neutron scattering cross-section, which is only sensitive to magnetic fluctuations perpendicular to $\mathbf{Q}$. The longitudinal character of magnetic excitations in Yb$_2$Pt$_2$Pb is a direct consequence of the strong orbital anisotropy imposed by the crystal field and the resulting strongly anisotropic Landé $g$-factor. There are profound and unusual consequences. It follows that even if the effective spin Hamiltonian that describes the low energy dynamics in $Yb_2Pt_2Pb$ has modes involving transverse spin fluctuations, such as spin waves, then they virtually do not couple to physical fields at our disposal and are *de facto* invisible in experiments. In particular, the measured longitudinal spectrum, which is typical of a spin-1/2 XXZ chain [Fig. 2], indicates the presence of transverse spinon excitations [9, 12–14], but these are not seen. This unobservable XY-part of the effective spin Hamiltonian (1), which in $Yb_2Pt_2Pb$ results from the well understood effect of quantum selection rules, may perhaps provide a useful insight into the nature of the cosmological dark matter phenomenon. The direct effect on our measurements is that we do not observe a (transverse) magnon, which is expected [15] when a magnetic field B = 4 T applied along (1-10) crystal direction saturates Yb moments that are parallel to the field [6–8], bringing this sublattice to the ferromagnetic (FM) state [Fig. 1(B)]. Instead, FM chains do not contribute to magnetic scattering, and this allows us to use the 4 T data as a background that can isolate their contribution at B = 0 [Fig. 2(D)].

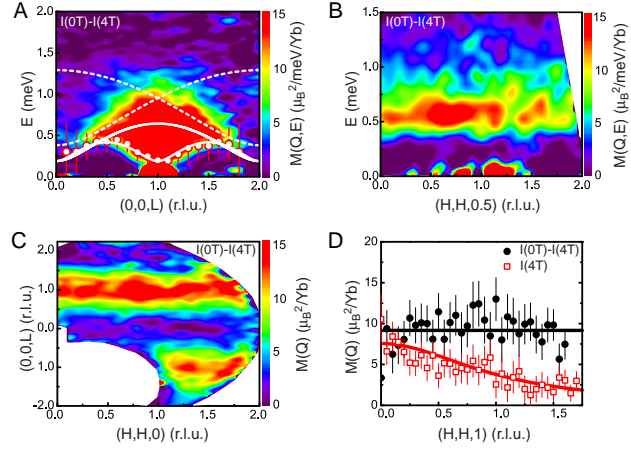

FIG. 2. **Fractional Spinon Excitations in Yb$_2$Pt$_2$Pb.** (A) The dispersion of the spectrum of magnetic excitations along the $Q_L$ direction in reciprocal space of Yb$_2$Pt$_2$Pb at T = 0.1 K, obtained by averaging the scattered neutron intensity over the first Brillouin zone in $Q_{HH}$, along the perpendicular, $(H, H, 0)$ direction. The circles mark the onset of the excitation continuum determined by fitting the constant-$Q_L$ data, while the solid white lines indicate the lower and upper boundaries of the 2-spinon continua, as described in the text. The broken lines are the upper boundaries of the 4-spinon continua (Supplementary Theory). (B) The dispersion of the scattered neutron intensity along the $Q_{HH}$ direction for $Q_L = 0.5 \pm 0.1$. (C) The partial static structure factor, M($\mathbf{Q}$), obtained by integrating the scattered intensity from 0.15 to 1.5 meV. M($\mathbf{Q}$) depends on the relative orientation of the scattering vector $\mathbf{Q} = (H, H, L)$ and the direction of magnetic moment fluctuations [11]. (D) Fluctuations along the magnetic field direction, (1-10), are perpendicular to the $(H, H, L)$ scattering plane and therefore yield intensity which is independent of the wave vector orientation in this plane. Such contribution is absent at 4 T, where the (1-10) sublattice is fully saturated (red points). The polarization factor, $P(4\text{T}) = \frac{(Lc^*)^2}{2(Ha^*)^2 + (Lc^*)^2}$, reflects the projection of the (110) sublattice moments on the scattering wave vector (red line), indicating that only magnetic fluctuations along the (110) moments, which are insensitive to magnetic field, contribute to magnetic scattering at 4 T. Both polarizations are present when $B = 0$ T, consistent with equal longitudinal spinon spectral weight in both (110) and (1-10) sublattices. By subtracting the 4 T data from the 0 T data, we can isolate the scattering from fluctuations along the field direction, (1-10), which are always perpendicular to the scattering vector $\mathbf{Q}$, and whose polarization factor is constant (black points). The good agreement between the polarization factor and our data confirms that only fluctuations of the Yb moments that lie along the respective, (110) or (1-10) direction are seen in our experiment, and that there is no measurable transverse component of magnetic moment or excitations. Error bars in all figures represent one standard deviation.

Having established the presence of a longitudinal spinon continuum in $Yb_2Pt_2Pb$ we now proceed to more quantitative analysis of the measured $M(\mathbf{Q}, E)$ lineshapes that aims to establish the hierarchy of energy scales in the effective $S = 1/2$ XXZ Hamiltonian. To this end, we fit the energy cuts at different values of $Q_L$ to a phenomenological half-Lorentzian line shape [16], which accounts both for the sharp continuum onset and its broad, asymmetric extent to higher energies (Fig. 3(A,B) and Supplementary Neutron Scattering Data). The obtained fits show $\chi^2$ near 1, which means that within the statistical accuracy of our data these *ad hoc* line shapes are indistinguishable from the measured $M(\mathbf{Q}, E)$. We can thus very accurately determine the lower boundary, $E_L(Q_L)$, of the spinon continuum (points in Fig. 2A), which we fit to the exact Bethe-Ansatz expression for the XXZ Hamiltonian (1) [9, 12, 13],

$$E_L(Q_L) = min\left\{\Delta_s + \varepsilon_s(Q_L), 2\varepsilon_s(Q_L/2)\right\}, \quad \varepsilon_s(Q_L) = \sqrt{I^2 \sin^2\left(\pi Q_L\right) + \Delta_s^2 \cos^2\left(\pi Q_L\right)}.(2)$$

Here $\Delta_s$ is the gap and $I$ the bandwidth of the spinon dispersion, $\varepsilon_s(Q_L)$, which both are functions of the J and $\Delta$ parameters of the Hamiltonian (1) (Supplementary Theory). The fit yields values $I = 0.322(20)$ meV and $\Delta_s = 0.095(10)$ meV for the spinon dispersion parameters, which correspond to $\Delta \approx 3.46$, and J = 0.116(10) meV in the effective spin-1/2 XXZ Hamiltonian, and the excitation gap at $Q_L= 1$, $E_g = 2\Delta_s = 0.19(2)$ meV. Despite the strong anisotropy of the individual Yb moments, their inferred coupling in the spin chain is surprisingly close to the isotropic Heisenberg limit $\Delta = 1$, as is evidenced by the smallness of the excitation gap $E_g$ compared to their observed bandwidth $\gtrsim 1$ meV [Fig. 2(A)].

Computations carried out on the XXZ Hamiltonian (1) closely reproduce key aspects of our experimental results. The mixed Heisenberg-Ising character of Yb$_2$Pt$_2$Pb is evident in the broad peak at $Q_L = 1$ in the structure factor M(**Q**) found by energy integrating the experimental and computed spectra [Fig. 3(C)]. M(**Q**) is intermediate between the near divergence expected for isotropic interactions ($\Delta = 1$) and the leading-order Ising expression [17] M(**Q**) $= 2$M$_0^2 \frac{1}{\Delta^2} \sin^2(\frac{\pi Q_L}{2})$, where M$_0 = \frac{1}{2} g_\parallel^{eff} \mu_B$ is Yb magnetic moment, $g_\parallel^{eff}$ being the effective spin-1/2 $g-$factor for the local Ising direction. Crystal electric field calculations for the Yb ground state doublet in Yb$_2$Pt$_2$Pb indicate $g_\parallel^{eff} = 7.9$ and $g_\perp^{eff} \lesssim 0.8$ (Supplementary Information), so that magnetic neutron scattering intensity, which is proportional to $\left(g_{\parallel,\perp}^{eff}\right)^2$, is at least 100 times weaker for the transverse, XY-polarized fluctuations, in agreement with what we observe.

Since we have been able to express the measured intensity in absolute units (Materials and Methods), the Q-integrated scattering in Fig. 3(C) yields a fluctuating moment $M_{fluct}^2 \approx 7.1 \mu_B^2$/Yb at 0.1 K, about half as large as the ordered moment $M_{order}^2$ determined in previous work [18]. The energy integral of the local autocorrelation function M($E$), which is obtained by integrating the measured intensity in Q, yields a similar result, $\approx 7.6 \mu_B^2$/Yb [Fig. 3(D)], with the difference indicating a systematic error resulting from different data binning. The sum rule for the effective spin-1/2 dictates that the integral intensity in each polarization channel is $(S^\alpha)^2 = 1/4$ ($\alpha = x, y, z$). Therefore, the sum $M_{fluct}^2 + M_{order}^2$ gives a total Yb moment, $M_{total}^2 = \left(\frac{1}{2} g^{eff} \mu_B\right)^2$. Combining the inelastic spectrum and the elastic order parameter measurements in Yb$_2$Pt$_2$Pb (Supplementary Neutron Scattering Data) we find $M_{total}$ that is between 3.8 and 4.4 $\mu_B/Yb$ [$g^{eff} = 8.2(5)$] for temperatures from 0.1 to 100 K [Fig. 4(A)], fully consistent with the predictions of the point charge model. The spinons provide virtually all of the magnetic dynamics in Yb$_2$Pt$_2$Pb, and they are completely captured by our experiments. This result immediately rules out a naive explanation that the observed longitudinal magnetic response could originate from the two-magnon continuum, as in conventional magnets, because in that case continuum would comprise only a small part of the dynamical spectral weight [19, 20]. Moreover, stable magnons do not exist in a spin-1/2 chain, where the elementary excitations are spinons, and the system's one-dimensionality is clearly established by the measured dispersion [Fig. 2(A-C)]. What is more, the static spin susceptibility $\chi_s$(T) computed for spin-1/2 XXZ chain with $g^{eff} \approx 7$ closely reproduces direct measurements of $\chi(T)$ [Fig. 4(B)].

Further comparison with the exact result [21] for the XXZ model (1), however, indicates that the fluctuations measured in $Yb_2Pt_2Pb$ at 0.1 K are stronger than the predicted spinon contribution to the dynamical spin structure factor, which for $\Delta = 3.46$ is only $\approx 20\%$ of the ordered spin contribution $S^2_{order}$. Fig. 3(D) makes it clear that the calculated $M(E)$ underestimates the contribution of the high-energy states in $Yb_2Pt_2Pb$. Direct comparison of the detailed energy dependencies of the measured [Fig. 2(A)] and computed (broadened by the instrumental resolution of 0.1 meV) [Fig. 4(A)] spectra of longitudinal excitations reveals that there is considerable spectral weight present in the experimental data above the upper boundary of the two-spinon continuum, $E_U(Q_L) = max\{\Delta_s + \varepsilon_s(Q_L), 2\varepsilon_s(Q_L/2)\}$, that is absent in the computed spectrum [12, 13]. A somewhat better agreement can be obtained by fitting the measured intensity to the calculated longitudinal structure factor $M(\mathbf{Q}, E)$ and adjusting $\Delta$ and J as fit parameters instead of adopting the values determined from the lower boundary of the continuum. This results in $\Delta = 2.6$ and J = 0.205 meV [Fig. 4(B)], shifting the two-spinon spectral weight to higher energy and also providing better agreement with the measured susceptibility [Fig. 4(D)] and M(Q) [Fig. 3(C)]. However, this improvement is achieved at the cost of the excellent experimental and theoretical agreement for the lower spinon boundary, which, in fact, is determined very precisely from the line fits [Fig. 3(A,B)]. This dilemma is resolved by noting that the observed high energy magnetic spectral weight in $Yb_2Pt_2Pb$ is consistent with a substantial contribution of four-spinon states, whose upper boundaries [14] are shown by the broken lines in Fig. 2(A). This result is quite unexpected, given that two-spinon excitations account for all but a few percent of the total spectral weight [14, 15, 22] in the nearest-neighbor Heisenberg-Ising chain.

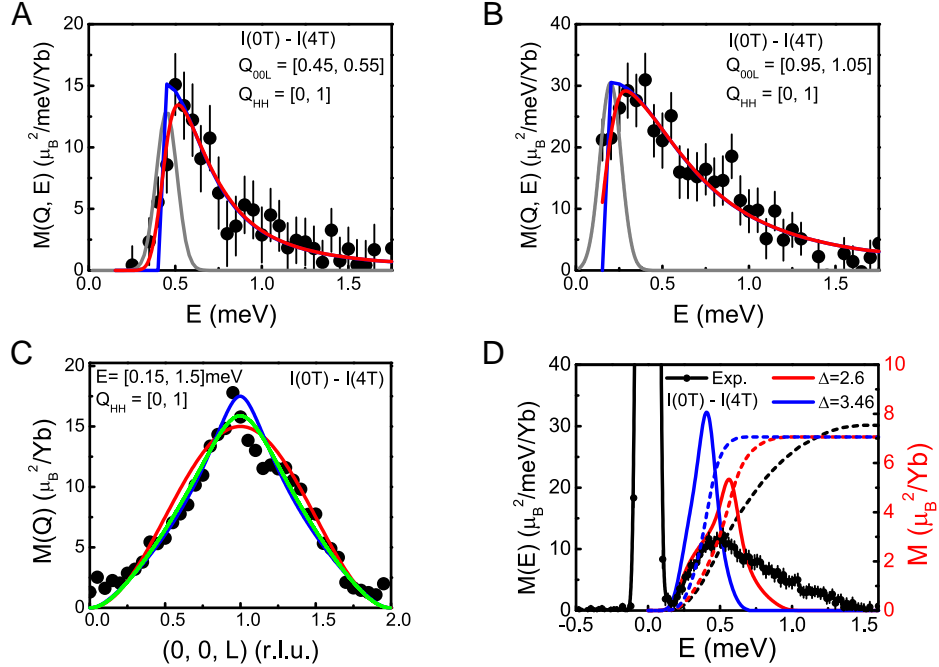

FIG. 3. **Spinon Lineshapes and the Onset of the Continuum in Yb$_2$Pt$_2$Pb.** The spectrum of the dynamical structure factor of magnetization fluctuations, M(**Q**, $E$), in Yb$_2$Pt$_2$Pb for (A) $Q_L = 0.50(5)$ ($q_{chain} = \pi/2$) and (B) $Q_L = 1.00(5)$ ($q_{chain} = \pi$), both integrated within 1 Brillouin zone in $Q_{HH}$. The red lines show fits to the "half-Lorentzian" lineshape (blue) convoluted with the Gaussian of 0.1 meV full width at half maximum (FWHM) representing the resolution of the DCS spectrometer (light grey), which provide a useful phenomenological way of quantifying the integral intensity and the onset of the continuum [16]. (C) The energy integrated scattering function M(**Q**) obtained by summing the normalized data over the first Brillouin zone in $Q_{HH}$ compares favorably with that calculated for the effective $S = 1/2$ Heisenberg-Ising Hamiltonian with $\Delta = 2.6$ and J = 0.205 meV and with the effective $g$−factor $g^{eff} \approx 10$ (blue line), and with $\Delta = 3.46$, J = 0.116 meV, and $g^{eff} \approx 13$ (green line). A fit to the leading Ising-limit ($\Delta \gg 1$) expression, M(**Q**) $= \frac{(g^{eff}\mu_B)^2}{2\Delta^2} \sin^2(\frac{\pi Q_L}{2})$, (red line) is less satisfactory, emphasizing that effective spin-1/2 Hamiltonian in Yb$_2$Pt$_2$Pb is not extremely Ising-like. This is consistent with the observation that the gap in the spinon spectrum, $E_g \approx 0.19$ meV (B), is markedly smaller than the bandwidth, $\gtrsim 1$ meV, Fig. 2A. (D) The energy dependence of the **Q**-integrated intensity, which represents the local dynamical structure factor, M($E$), of magnetization fluctuations in Yb$_2$Pt$_2$Pb. The energy-integral of the M($E$) inelastic intensity (black dashed line, right scale) gives the square of the total fluctuating magnetic moment of $\approx 7.6\mu_B^2$ per Yb. Computational results for M($E$) and its energy integral are compared for $\Delta = 2.6$ (red solid and dashed lines) and $\Delta = 3.46$ (blue solid and dashed lines).

It turns out that all these remarkable and seemingly perplexing experimental results have a natural and even elegant theoretical explanation, if one steps back and considers the 4f-electron exchange in $Yb_2Pt_2Pb$ in the presence of strong spin-orbit coupling and a crystal field that lifts the large orbital degeneracy of the $J = 7/2$ multiplet. The inter-site electron hopping in the f-electron Hamiltonian for $Yb_2Pt_2Pb$, which we adopt in the form of a one-dimensional Hubbard model (Supplementary Theory), leads to a new electronic interaction [23], whose physical nature is not a Heisenberg-Dirac spin-exchange [1, 2], but rather an orbital-exchange (Fig. 1), a realization that has been appreciated in the physics of Kondo effect [24, 25] and more recently in certain cold atom systems [26]. A reformulation of the exchange coupling that takes into account orbital character [27] has just the features that are needed to explain our remarkable results in $Yb_2Pt_2Pb$.

The orbital-exchange interaction in $Yb_2Pt_2Pb$ is a natural generalization of the Heisenberg-Dirac spin-exchange between the two electrons, and has the same physical origin in the electronic

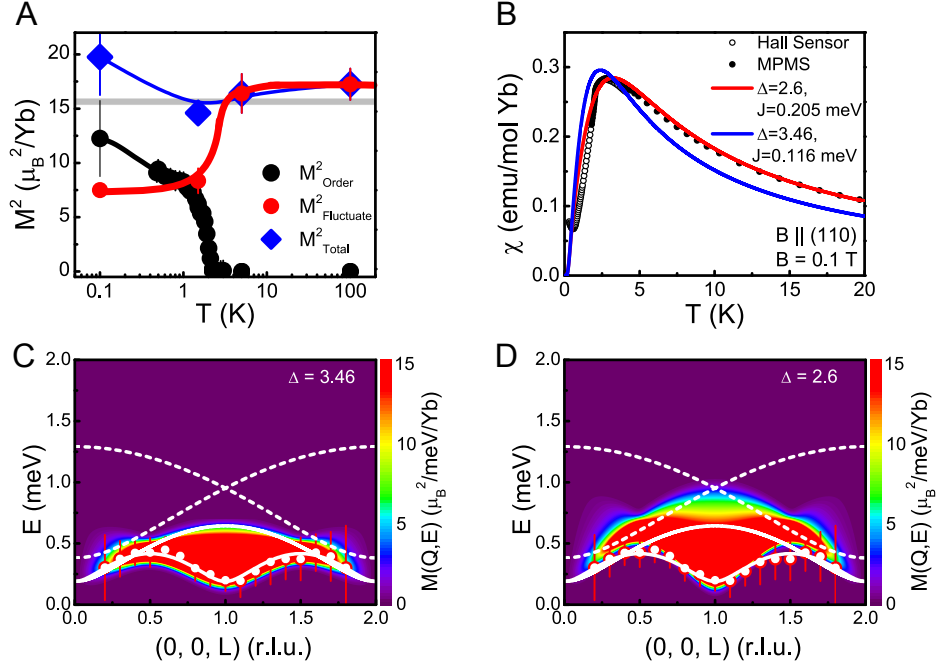

FIG. 4. **Spinons in Yb$_2$Pt$_2$Pb: Theory and Experiment.** (A) Temperature dependencies of the ordered Yb moment from neutron diffraction measurements (black circles), the fluctuating moment from the energy and wave vector integrated normalized M(Q,E) (red points), and the total (blue points). (B) The temperature dependence of the static, uniform magnetic susceptibility $\chi(T)$ for Yb$_2$Pt$_2$Pb (black circles), measured with a Magnetic Properties Measurement System ($T \geq 1.8$ K) and a Hall sensor magnetometer (0.2 K$\leq T \leq$ 4 K), shows good agreement with $\chi(T)$ calculated for the XXZ chain, for $\Delta = 2.6$ and J = 0.205 meV (red line, $g^{eff} \approx 7.4$). Agreement is less good for $\Delta = 3.46$ and J = 0.116 meV (blue line, $g^{eff} \approx 6.5$). (C), (D) The longitudinal structure factor, M($Q, E$), of the XXZ spin-1/2 chain (1) calculated using the algebraic Bethe ansatz [13, 14, 22] (C) for $\Delta = 3.46$ and J = 0.116 meV, and (D) for $\Delta = 2.6$ and J = 0.205 meV. The experimentally determined lower boundary $E_L(Q_L)$ is shown (circles) along with the calculated lower and upper 2 spinon (solid lines) and the upper 4 spinon (broken lines) boundaries, as described in the text. The calculation is normalized to the total experimental intensity by using the sum rule for a single component of the dynamical structure factor, which holds for spin-1/2, $\iint S^{zz}(Q, E)\frac{dQ}{2\pi}dE = \frac{1}{4}$.

Coulomb repulsion [1, 2]. Underlying it is a simple physical picture schematically illustrated in Fig. 1. The magnetism in Yb$_2$Pt$_2$Pb is tied to the wave function of a single 4f hole with orbital momentum $L = 3, |m_L| = 3$, having six-fold symmetry around the $\boldsymbol{J}$ quantization axis, given by the magnetic structure as perpendicular to the rails of Yb ladders in Yb$_2$Pt$_2$Pb crystal. As in the more familiar case of spin exchange, the energy cost for hopping between sites, which in Yb$_2$Pt$_2$Pb is synonymous with orbital exchange, is reduced when neighboring Yb ions are in alternating states of $m_L = \pm 3$, since here the exchange of electrons between the two sites required for hopping involves the overlap of two identical orbital lobes along the ladder rails [Fig. 1(A,C)]. The six-fold symmetry of the f-orbital breaks the leg-rung equivalence and ensures that this energy advantage is not accrued for hopping in a transverse direction, decoupling the ladder legs. Combined with the weak interactions between orthogonal ladders mandated by the SSL geometry [5] this leads to the spin-chain nature of the emergent effective Hamiltonian.

In technical terms (Supplementary Theory), the leading-order Coulomb contribution for the low-energy manifold of electronic states is given by the two-electron permutation operator, $P_{12}$, which in the cases where only electronic spins are at play reduces to the usual Heisenberg spin-exchange, $\sim \mathrm{J}\boldsymbol{S}_1\boldsymbol{S}_2$. For the case of a $\boldsymbol{J}$-manifold, which in the absence of crystal fields is highly degenerate, it has the form of a permutation operator acting on a $(2J+1) \times (2J+1)$-dimensional space of two neighboring Yb ions. The permutation operator interchanges states $|m_{J1}, m_{J2}\rangle$ and $|m_{J2}, m_{J1}\rangle$ with equal weights, thus including the process $|7/2, -7/2\rangle$ to $|-7/2, 7/2\rangle$ where both moments simultaneously reverse, states that are uncoupled by conventional Heisenberg-Dirac spin-exchange, Fig. 1(C,D). The crystal field lifts the degeneracy of the Yb moments, and while the effective interaction that emerges after the projection on the manifold of the lowest Kramers doublets $m_J = \pm 7/2$ has the form of the antiferromagnetic $S = 1/2$ XXZ Hamiltonian, the parent Hamiltonian for all spin chain systems including $Yb_2Pt_2Pb$, it retains the birthmark of its unusual origin in exchange processes that are distinct from those having the conventional Heisenberg $\boldsymbol{J}_1\boldsymbol{J}_2$ form.

The effective spin-1/2 physics emerges in $Yb_2Pt_2Pb$ from the combination of high-energy (Coulomb, spin-orbit, hopping) interactions much as high-temperature superconductivity in cupric oxides is believed to emerge from electronic interactions governing the Mott insulating state, which similarly act on atomic energy scales, orders of magnitude larger than the highest temperatures relevant for superconductivity [28]. In $Yb_2Pt_2Pb$ the spin-orbit coupling virtually quenches the electronic spin degree of freedom, forcing its alignment with the large orbital moment, and in this way the effective spin-1/2 XXZ model effectively describes the quantum dynamics of the electronic orbital degree of freedom. This is directly evidenced in our experiments by the large, $\approx 4\mu_B$ magnetic moment carried by spinons. The orbital exchange sets the scale for these emergent quantum dynamics, which we find by comparing the measured spinon dispersion with computed spectra [Fig. 4].

Since the orbital angular momentum dominates the total Yb moment, magnetic order in $Yb_2Pt_2Pb$ is synonymous with orbital order, and the configuration depicted in Fig. 1(D,E) is a natural way to understand how permutation of two neighboring electrons generates two spinons in the antiferromagnetic background. This is a process that entails charge-orbital separation, since the electron count per site is unchanged by correlated hopping, while the phases of the orbital wave function on both sites are reversed. Further-neighbor orbital exchange leads to states with four spinons [Fig. 1(E)]. Hence, long-range hopping, either by virtue of the in-chain itinerancy of the 4f electrons, or via coupling to the conduction electrons in metallic $Yb_2Pt_2Pb$, provides a natural mechanism for the spectral weight of the excitations that we observe above the 2-spinon but within the 4-spinon continuum boundaries.

By conventional wisdom, one could hardly imagine a less likely place to find fractional quantum spinon excitations than in metallic rare-earth based systems, where strong spin-orbit coupling leads to large, classical moments with strong anisotropy [29]. Our results turn this consensus on its head, showing definitively that quantum magnetic dynamics in $Yb_2Pt_2Pb$ involve the reversal of large and orbitally dominated Yb moments, opening a veritable Pandora's box of unique novel phenomena that are similarly unexpected. We have demonstrated unprecedented experimental access that is selective to the longitudinal excitations of the XXZ model, offering unique opportunities for comparisons with theory, and in particular, our measurements challenge theories of integrable models to include the effects of longer-range interactions. The low-energy effective spin-1/2 Hamiltonian that emerges from orbital exchange has the remarkable feature that its transverse excitations, while contributing to the energy and entropy of the system, do not couple to magnetic fields. Similarly invisible is the transverse spin wave in the ferromagnetic phase at high-field, a "dark magnon" quasi-particle, which perhaps offers an insight into the nature of the weakly interacting massive particles (WIMPs) of the cosmological dark matter. Finally, our results provide a specific mechanism for charge-orbital separation in $Yb_2Pt_2Pb$, where the proliferation of spinons implies that electrons lose their orbital-phase identity. When united with the previous demonstrations of spin-charge and spin-orbital separation, this finding completes the triad of electron fractionalization phenomena in one dimension [30–32].

## ACKNOWLEDGMENTS

Work at Brookhaven National Laboratory (I.A.Z., A.M.T., M.S.K.) was supported by the Office of Basic Energy Sciences (BES), Division of Materials Sciences and Engineering, U.S. Depart-ment of Energy (DOE), under contract DE-SC00112704. Work at Stony Brook (L.S.W., W.J.G., M.C.A.) was supported by NSF-DMR-131008. L.S.W. was also supported by the Laboratory Directed Research and Development Program of Oak Ridge National Laboratory (ORNL). This research at ORNLs Spallation Neutron Source was sponsored by the Scientific User Facilities Division, Office of Basic Energy Sciences, U.S. DOE. Work at NIST Center for Neutron Research (NCNR) is supported in part by the NSF under Agreement no. DMR-1508249. J.-S.C. and M.B. acknowledge support from the Netherlands Organization for Scientific Research (NWO) and the Foundation for Fundamental Research on Matter (FOM) of the Netherlands.

---

# Supplementary Materials

## Orbital-Exchange and Fractional Quantum Number Excitations in an f-electron Metal Yb$_2$Pt$_2$Pb

L. S. Wu,[1,2,3] W. J. Gannon,[1,2,4] I. A. Zaliznyak,[2,*] A. M. Tsvelik,[2] M. Brockmann,[5,6] J.-S. Caux,[6]

M. S. Kim,[2] Y. Qiu,[7] J. R. D. Copley,[7] G. Ehlers,[3] A. Podlesnyak,[3] and M. C. Aronson[1,2,4]

[1]*Department of Physics and Astronomy, Stony Brook University, Stony Brook, NY 11794, USA*

[2]*Condensed Matter Physics and Materials Science Division,*

*Brookhaven National Laboratory, Upton, NY 11973, USA*

[3]*Quantum Condensed Matter Division, Oak Ridge National Laboratory, Oak Ridge, TN 37831, USA*

[4]*Department of Physics and Astronomy, Texas A&M University, College Station, TX 77843, USA*

[5]*Max Planck Institute for the Physics of Complex Systems,*

*Nöthnitzer Str. 38, 01187 Dresden, Germany*

[6]*Institute for Theoretical Physics, University of Amsterdam,*

*Science Park 904, 1098 XH Amsterdam, The Netherlands*

[7]*NIST Center for Neutron Research, National Institute of Standards and Technology, Gaithersburg, MD 20899 USA*

(Dated: June 1, 2016)

**This PDF file includes:**

Materials and Methods

Supplementary Text:

Crystal electric field splitting of the Yb $J = 7/2$ multiplet and the single ion anisotropy

Supplementary Theory

Supplementary Neutron Scattering Data

Tables S1 to S3

Figures S1 to S10

Supplementary References [31] to [53]

---------

* Correspondence to: zaliznyak@bnl.gov

# I.  MATERIALS AND METHODS

## A.  Methods summary

Inelastic neutron scattering measurements on $Yb_2Pt_2Pb$ were carried out using the Disk Chopper Spectrometer (DCS) [31] at the Center for Neutron Research at the National Institute of Standards and Technology and the Cold Neutron Chopper Spectrometer (CNCS) [32] at the Spallation Neutron Source at Oak Ridge National Laboratory.  A 6 g sample of $\approx 400$ co-aligned $Yb_2Pt_2Pb$ crystals was used, oriented with (1-10) crystal lattice direction vertical and the $(Q_{HH},Q_{HH},Q_L)$ reciprocal lattice plane horizontal.  On DCS the sample was mounted in a 10 T vertical field superconducting magnet equipped with the dilution refrigerator, on CNCS a 5T vertical field magnet with 1.5K base temperature was used.  The incident neutron energy was $E_i = 3.27$ meV ($\lambda = 5.0$ Å) on DCS and $E_i = 3.316$ meV ($\lambda = 4.97$ Å) on CNCS. For nominal zero field measurements at base temperature 0.07 K a small bias field of about 250 Oe was used to suppress the superconductivity of the aluminum sample holder.  A 4 T field along the (1-10) direction fully polarizes the sublattice with magnetic moments parallel to (1-10), allowing us to use measurements at 4 T as a background to isolate the inelastic scattering from the orthogonal sublattice with moments along (110), which is unaffected by the field. The measured intensities were corrected for neutron absorption and the scattering cross section was put on an absolute scale by comparing the measured Bragg peak intensities with calculated nuclear structure factors. All wave vectors, $(Q_{HH},Q_{HH},Q_L)$, are indexed in reciprocal lattice units of $a^* = 2\pi/a = 0.81$ Å$^{-1}$ and $c^* = 2\pi/c = 0.91$ Å$^{-1}$.  The susceptibility of XXZ model (S18) was calculated using the quantum transfer matrix approach suitable for finite temperature calculations [33, 34] and solving the corresponding non-linear integral equations [35].

## B.  Sample Preparation

$Yb_2Pt_2Pb$ single crystals were grown from a Pb flux as described in Ref. 5.  The typical crystals have rod-like shape with typical dimensions $\sim 0.5$ mm in width and $\sim 4-5$ mm in length, and an approximately square cross-section with flat surfaces normal to the (110) crystal direction and with the (001) lattice $c-$axis along the length.  Crystal orientations were checked with X-ray diffraction. The $\approx 6$ g single crystal sample used for inelastic neutron scattering was assembled by co-aligning, using the crystal morphology, between 300 and 400 crystals on 6 aluminum plates.  Crystals were held in place by the surface tension from hydrogen free Fomblin oil.  These plates were clamped together in an Al box, resulting in the overall crystal mosaic spread perpendicular to the (001) direction of $\approx 2°$.  A picture of the sample holder is shown in Fig. S1A. Neutron powder diffraction measurements were performed on a 5 g powder sample prepared from similar single crystals.

## C. Magnetization measurements

For T= 1.5 K and T> 1.8 K, the bulk magnetization and susceptibility were measured on a single crystal sample using a [Quantum Design] Magnetic Property Measurement System SQUID magnetometer with a gas-flow cryostat. For T< 1.8 K, magnetization and susceptibility were measured on a single crystal using a Hall sensor magnetometer, mounted in the dilution refrigerator insert of the [Quantum Design] Physical Property Measurement System.

## D. Neutron Scattering setups

Most of the data shown in the main manuscript were acquired using the Disk Chopper Spectrometer (DCS) at the National Institute for Standards and Technology (NIST) Center for Neutron Research (NCNR) [31, 36]. Additional measurements of the temperature dependence were performed using the Cold Neutron Chopper Spectrometer (CNCS) at the Spallation Neutron Source at Oak Ridge National Laboratory [32, 37]. The complementary neutron powder diffraction measurements quantifying the order parameter and its detailed temperature dependence were made on the BT-7 double focusing triple-axis spectrometer at NIST NCNR [38] and are reported elsewhere [18].

For the single crystal measurements on DCS and CNCS, the sample holder with the sample was mounted in the $Q_{HH} - Q_L$ scattering plane, with the (1-10) direction vertical. For the DCS experiment, a 10 T cryomagnet with a dilution refrigerator insert with base temperature of 0.1 K was used. A 1.5 cm × 2.5 cm beam mask was installed in front of the sample to reduce the unwanted background scattering from the sample environment. A small bias field of 250 Oe was used for the nominal zero field measurements at base temperature in order to suppress the superconductivity of the aluminum sample holder. The magnet dark angle was set at 75° from the (110) direction, eliminating a substantial source of background scattering. The choppers were set to select neutrons with fixed incident energy of $E_i$ = 3.27 meV ($\lambda_i$ = 5.0 Å). The vertical acceptance of the detector system to scattered neutrons is limited by the height of the magnet window. Only neutrons scattered from the sample within angle ±θ and illuminating a segment of height $H$ on the detector are counted, as illustrated in Fig. S1B. On DCS the distance from sample to detector is $D$ = 4010 mm, and only the data of the central detector bank with tubes of height $H$ = 400 mm [36] were analyzed, thus resulting in the vertical acceptance of ±$\theta_{DCS} \approx \pm 5.7°$ in our experiment. As a result, although the scattering plane is (HHL), part of the scattering from the (H-HL) plane within this angular acceptance also contributes to the overall scattering intensity. The corresponding vertical wave vector resolution full width at half maximum (FWHM) is $\approx 0.25$ Å, or about 0.2 reciprocal lattice unit of $a_{110}^* = a^*\sqrt{2} \approx 1.146$Å$^{-1}$.

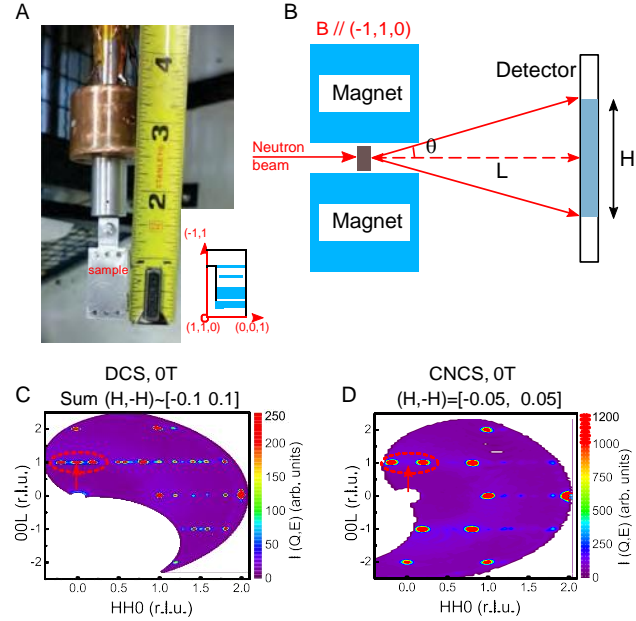

FIG. S1. (A) Single crystal sample holder used for neutron scattering experiments at DCS at NCNR, NIST, and CNCS, SNS. Single crystals of $Yb_2Pt_2Pb$ were aligned with the (110) crystal direction normal to the surface of the holder plate, and with the $(1-10)$ direction parallel to the long side of the rectangular holder. (B) Schematic view of the configuration for both DCS and CNCS neutron scattering setups. Magnetic field B is applied along the vertical, $(1-10)$ direction. The sample rotates in the horizontal, (H H L) plane, with access to the vertical (H -H L) scattering plane limited by the acceptance angle, $\pm\theta$. The distance $D$ from sample to detector and the detector height, $H$, for both DCS and CNCS experiments are described in the text. (C), (D) Contour maps of elastic neutron scattering intensity integrated within the energy range $E = [-0.2, 0.2]$ meV, measured in zero field on DCS (C) and CNCS (D). For the DCS data, the detectors are only position sensitive in the horizontal direction, and all scattering in the vertical, (H -H) direction is projected onto the (H H L) scattering plane, (C). For the CNCS data, the detectors have spatial resolution in both horizontal and vertical directions. The data in (D) are integrated only within a narrow region, $Q_{H-H} = [-0.05, 0.05]$, in the vertical direction. The diffraction seen in the DCS data at $Q = (001)$ (C) is picked up by the relaxed vertical resolution of that measurement and is absent in the CNCS data (D), as indicated by red arrows.

For the CNCS experiment, a 5 T cryomagnet with a He flow cryostat with T= 1.5 K base temperature was used. The horizontal angular acceptance of the magnet was large and was not significantly restricted by the magnet dark angle. The incident neutron energy was fixed at 3.316 meV ($\lambda_i = 4.97$ Å), and the high-flux mode of the instrument was used to maximize neutron intensity. The distance from sample to detector was $D = 3500$ mm. Although each 128-pixel detector tube has height $H = 2000$ mm [32, 37], the vertical acceptance of the magnet constrained us to only use the middle 88 pixels in the vertical direction.

This gives an effective height of the detector in this experiment of $H = 2000 \times 88/128 = 1375$ mm, and an angular acceptance $\pm\theta_{\text{CNCS}} \simeq \pm 11°$, nearly twice the vertical acceptance at DCS. This somewhat increased vertical acceptance angle makes difference for the inelastic neutron scattering measurements by virtue of increasing the overall scattering intensity, but it is of even greater importance for the elastic diffraction measurements. Here, Bragg reflections that are close to, but not exactly in the horizontal scattering plane, may appear unexpectedly if one simply adds all scattered intensity along the vertical dimension of the detectors, thus yielding a misleading result. Fortunately, the CNCS detectors are position sensitive along the vertical direction [32, 37], which provides an access to the out of plane, (H-HL) direction and allows isolating contributions from the (HHL) scattering plane.

By comparing the neutron intensity measured on the DCS with that measured on the CNCS within different vertical wave vector ranges, we conclude that the DCS measurements, which are not position sensitive in the vertical direction, correspond to summing up the intensity over the range of $Q_{H-H} = [-0.1, 0.1]$, in agreement with the FWHM of the vertical resolution of the DCS measurements estimated above. Contour maps of the elastic ($E = [-0.2, 0.2]$ meV) neutron scattering intensity measured at zero field on DCS and CNCS are shown in Fig. S1C and Fig. S1D, respectively. For the DCS data all scattering within the vertical acceptance is projected onto the (HHL) scattering plane. Thus, a diffraction peak is seen at $Q = (001)$ in Fig. S1C, which is actually tails of the satellite magnetic Bragg peaks in the vertical, (H-HL) plane, above and below the (HHL) scattering plane, which are picked up by the vertical resolution of this measurement. For the CNCS data we can limit the integration in the vertical direction to a narrow range, $Q_{H-H} = [-0.05, 0.05]$, thus avoiding the out-of-plane (H-HL) contributions. Consequently, the scattering at $Q = (001)$ is absent, as indicated by the red arrow in Fig. S1D. This will be further discussed later, with reference to Fig. S8A, B.

For powder neutron diffraction measurements on BT-7, a 5 g powder sample was pressed in an Al sample can and mounted in a $^3$He cryostat. Neutrons of wavelength $\lambda_i = 2.359$ Å ($E_i = 14.7$ meV) were incident on the sample, and scattered neutrons were measured using the instrument's position sensitive detector.

### E. Absolute normalization and extinction effect

Since our neutron scattering experiments were carried out on two different time of flight spectrometers, DCS and CNCS, in order to compare and combine these two data sets we express both in absolute units. In principle, we could carry out this normalization in several ways: by using the incoherent scattering of the sample, the incoherent scattering of a vanadium standard, or using the structure factor of the diffraction peaks of $Yb_2Pt_2Pb$. The incoherent scattering of Pt and Pb are very small, and the incoherent scattering of the $Yb_2Pt_2Pb$ sample itself is comparable to scattering from the sample environment, ruling out this approach. Since Yb and Pt have relatively large absorption cross sections $\sigma_{abs}$, combined with the fact that we have a relatively large sample, normalization using a vanadium standard is also problematic, due to this absorption effect. Consequently, we use the structure factors of nuclear and magnetic diffraction peaks in $Yb_2Pt_2Pb$ to place the experimental intensities on an absolute scale.

| Nuclear Diffraction | (002) | (110) |
|---|---|---|
| $|F_{hkl}|^2$ (barn·fu$^{-1}$) | 1.19927 | 5.5389 |
| $\int d^3Q \int dE I(Q,E)$ (meV) | 0.00165 | 0.00281 |

TABLE S1. The calculated nuclear structure factors and the experimentally measured integrated intensities for the primary diffraction peaks accessed in the (HHL) scattering plane in the CNCS single crystal neutron scattering measurements at 1.5 K in zero field.

| AF Diffraction | (1.2, 1.2, 0) | (1.8, 1.8, 0) | (0.2, 0.2, 1) | (0.8, 0.8, 1) |
|---|---|---|---|---|
| $|Q_{hkl}|$ (A$^{-1}$) | 1.373 | 2.06 | 0.9239 | 1.280 |
| $|F(Q)|^2$ | 0.924 | 0.84 | 0.964 | 0.933 |
| $|F_{hkl}|^2$ (barn·fu$^{-1}$) | 0.0636 | 0.1187 | 1.3379 | 2.725 |
| $\int d^3Q \int dE I(Q,E)$ (meV) | 0.0001219 | 0.00023148 | 0.001805 | 0.002754 |

TABLE S2. List of the calculated antiferromagnetic structure factors, $|F_{hkl}|^2$, and the experimentally measured integrated intensities for magnetic diffraction peaks accessed in the (HHL) scattering plane in the CNCS single crystal neutron measurements at 1.5 K in zero field ($|F(Q)|^2$ is magnetic form factor [39] squared). The ordered moment at 1.5 K is $M_{order} = 2.5\mu_B$/Yb, which was extracted from the order parameter measurement shown in Fig. S2A.

Using the known crystal structure [5], we have calculated the nuclear structure factors for the diffraction peaks accessed in the (HHL) scattering plane, two of which are listed in Table S1. Magnetic structure factors were calculated based on the powder diffraction experiments on BT-7. Measurements for scattering angles $15° \leq 2\theta \leq 90°$ were performed at different fixed temperatures (T = 10 K, 1.5 K, and 0.5 K) above and below the ordering temperature $T_N = 2.07$ K, and the zero field antiferromagnetic (AF) magnetic structure was solved [18]. Scattered intensity was measured as a function of temperature at $2\theta = 20°$, tracking the AF order parameter at $Q = (0.2, 0.2, 1)$ from 0.48 K up to 3 K, as shown in Fig. S2A. The magnetic transition is clearly seen at T$\approx$ 2.1 K, the Néel temperature previously identified in specific heat and magnetic susceptibility measurements [5, 6]. Based on the solution of the magnetic structure [18], and the fitted moment values from the $2\theta$ scans at fixed temperatures, it was possible to extract the ordered moment, $M_{order} = 2.5\mu_B$/Yb at 1.5 K, as indicated by the red solid diamond in Fig. S2A. The calculated

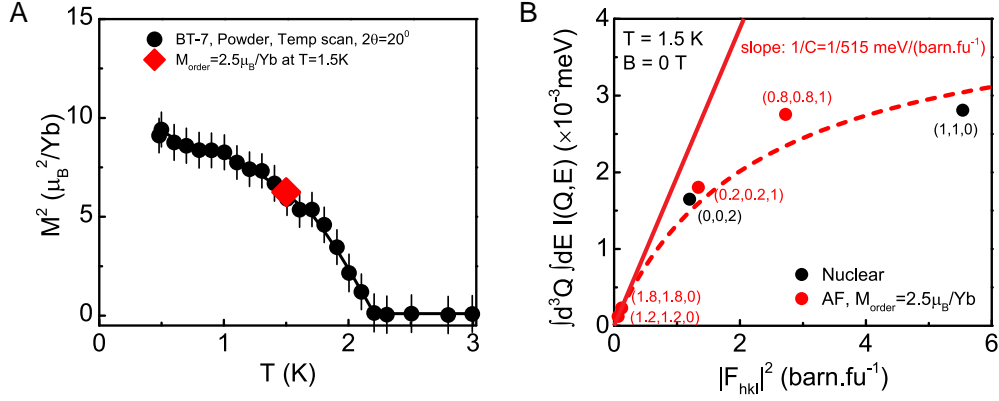

FIG. S2. **Magnetic and nuclear Bragg intensities in Yb$_2$Pt$_2$Pb in zero field.** (A) Ordered moment squared as a function of temperature. Black solid circles are the temperature scan at the (0.2, 0.2, 1) AF peak position at $2\theta = 20°$ [18], which was put on the absolute scale using the ordered moment $M_{order} = 2.5\mu_B/$Yb (red solid diamond) extracted from the refinement of the powder diffraction pattern measured in a $2\theta$ scan at 1.5 K. (B) Comparison of the calculated nuclear and magnetic structure factors and the experimental integrated peak intensities for peaks accessed at 1.5 K at CNCS and listed in Tables S1 and S2. The slope in the limit of small intensity gives the normalization factor for the experimental CNCS data to absolute values, $1/C = \int d^3Q \int dEI(Q,E)/|F_{hkl}|^2$ (solid red line). The dashed line is a fit to the semi-empirical secondary extinction factor for the single crystal diffraction intensity, as explained in the text.

magnetic structure factors for diffraction peaks in the (HHL) plane are listed in Table S2.

By comparing the calculated nuclear and magnetic structure factors, $|F_{hkl}|^2$, with the experimentally measured integrated intensities, $\int d^3Q \int dEI(Q,E)$, the normalization factor $C$ can be obtained as a ratio,

$$C = \frac{|F_{hkl}|^2}{\int d^3Q \int dEI(Q,E)}. \tag{S1}$$

However, if we plot the experimentally measured integrated intensity as a function of the calculated structure factor [Fig. S2B], the result does not fall on a straight line. This is due to the secondary extinction effect, which is usually significant for single crystal diffraction experiments on large mosaic crystals. To minimize this extinction effect, we selected the two weakest magnetic diffraction peaks, (1.2, 1.2, 0) and (1.8, 1.8, 0), to extract the normalization factor $C$. As indicated by the red solid line in Fig. S2B, the inverse of the slope gives the normalization factor, $C = 515$ barn·meV$^{-1}$. On the other hand, an approximate account for the secondary extinction in neutron diffraction can be accomplished by using the following exctinction-corrected expression for the measured integrated Bragg intensity as a function of the calculated structure factor,

$$\int d^3Q \int dEI(Q,E) = C\frac{|F_{hkl}|^2}{1 + \alpha|F_{hkl}|^2}, \tag{S2}$$

where $\alpha$ is determined by a combination of the sample mosaic and its effective thickness. The dashed line shows fit of our data to Eq. (S2) with $C = 515$ barn·meV$^{-1}$ and $\alpha = 0.44$. The fit provides a good description of the data, and we will use this normalization factor $C$ to express the experimental scattering intensity in absolute units of barn $\cdot$ meV$^{-1}$.

In discussing magnetic neutron scattering, it is illuminating to make a direct comparison of the fluctuating magnetic moments with static moments determined from magnetization measurements, so it is convenient to convert the measured cross-section to units of $\mu_B^2 \cdot$ meV$^{-1}$. The cross-section of (unpolarized) neutron magnetic scattering, which is proportional to the measured intensity, $I(Q,E)$, can be expressed as,

$$\frac{d^2\sigma(Q,E)}{dEd\Omega} = \frac{k_f}{k_i} \left(\frac{\gamma r_0}{2\mu_B}\right)^2 e^{-2W} \sum_\alpha \left(1 - \widetilde{Q}_\alpha^2\right) M^{\alpha\alpha}(Q,E) = \frac{k_f}{k_i} \cdot C \cdot I(Q,E), \tag{S3}$$

[40], where $k_f$ and $k_i$ are the scattered and the incident neutron wave vectors, $r_0 = 2.818 \times 10^{-13}$ cm is the classical electron radius, $\gamma = 1.913$ is the magnetic moment of the neutron in nuclear magnetons, $(\gamma r_0)^2 = 0.291 \times 10^{-24}$ cm$^2 = 0.291$ barn [11], $\alpha$ denotes the Cartesian coordinate, $x, y$, or $z$, which characterizes the polarization of magnetic fluctuations, and $\widetilde{Q}_\alpha = Q_\alpha/Q$ are the projections of the unit vector along the wave vector transfer direction on the Cartesian axes. $e^{-2W}$ is the Debye-Waller factor, which at low temperatures and small wave vectors of our measurements is taken to be unity. $M^{\alpha\alpha}(Q,E)$ is the correlation function of magnetic moments, which is related to the spin-spin correlation function, $S^{\alpha\alpha}(Q,E)$, computed for the effective spin-1/2 model through,

$$M^{\alpha\alpha}(Q,E) = (g_\alpha^{eff})^2 \mu_B^2 |F(Q)|^2 S^{\alpha\alpha}(Q,E), \tag{S4}$$

where $g_\alpha^{eff}$ is the effective $g-$factor, and $|F(Q)|$ is the magnetic form factor for Yb$^{3+}$ local moments [39]. Using Eqs. (S2)–(S4), we can express both the measured scattering intensity, $I(Q,E)$, and the calculated spin-spin correlation function, $S^{\alpha\alpha}(Q,E)$, in units of $\mu_B^2 \cdot$ meV$^{-1}$ [41].

The absolute cross section is based on the structure factor calculated for a crystal unit cell with eight Yb atoms. Thus, in the absence of magnetic field, the scattering must be normalized by a factor of eight on account of 8 Yb sites per unit cell (and also keeping track of the different sublattice polarizations) in order to express it in magnetic units of $\mu_B^2 \cdot$ meV$^{-1}$/Yb. As discussed above, strong Ising anisotropy divides Yb atoms into two orthogonal sublattices, with Ising axes along (110) and (-110) directions, respectively. Magnetic field of 4 T along (110), or (-110), polarizes moments in one of these two sublattices, so that only a single sublattice with four Yb moments per unit cell contributes to the measured magnetic scattering at high fields. Hence, in the 4 T data set, where one sublattice is completely polarized, the remaining inelastic scattering is from the orthogonal 4-Yb sublattice only, and must instead be normalized by a factor of four.

## F. Absorption Correction

As mentioned above, Yb and Pt have relatively large absorption cross sections (Table S3), and therefore account for neutron absorption in the sample is important for consistent comparison of the DCS and the CNCS data and for the absolute intensity normalization. Here we explain how absorption has been included in the normalization factor $C$. We estimate that the absorption correction dominates the systematic error, and that the comparisons of the DCS and CNCS data sets are accurate within $\approx 10\%$.

|    | $\sigma_{coh}$(barn) | $\sigma_{inc}$(barn) | $\sigma_{abs}$ (barn) |
|----|----------|----------|----------|
| Yb | 19.42    | 4        | 34.8     |
| Pt | 11.58    | 0.13     | 10.3     |
| Pb | 11.115   | 0.003    | 0.171    |

TABLE S3. Neutron scattering and absorption cross sections for Yb, Pt and Pb [42].

Although the dimensions of the aluminum sample holder are $\simeq 1.5$ cm wide, 2.5 cm high, and 0.8 cm thick, the overall effective shape of the horizontal cross-section of our multi-crystal $Yb_2Pt_2Pb$ sample with six co-aligned layers of single crystals can be approximated by a rectangle of width $\simeq 0.8$ cm along the crystal $c$ direction and thickness $\simeq 0.3$ cm along the crystal (110) direction, which is perpendicular to the sample holder surface.

For the DCS data measured at 0.1 K, the absorption correction has been carried out using the absorption option in the DCS Mslice program from the DAVE analysis software package [43]. In order to understand the magnitude of systematic errors associated with the absorption, we compare the data with and without the absorption correction in Fig. S3. The absorption effect is evident in the inelastic scattering data of Figure S3A, which shows the raw measured inelastic neutron intensity in the (HHL) scattering plane summed over the energy range $E = [0.15, 1.5]$ meV, without the absorption correction. The white dashed arcs in Fig. S3A indicate the region of suppressed intensity where the absorption is maximum. It occurs when neutron beam is parallel to the crystal $c-$axis, the long dimension of the crystal. This decrease in the intensity is absent once the absorption correction is applied, and the overall intensity is also enhanced [Fig. S3B]. The normalization procedure accounting for the extinction correction described earlier has been performed for the DCS data both without [Fig. S3C] and with [Fig. S3D] the absorption correction. The dashed line in Fig. S3C shows fit of the uncorrected data using Eq. (S2), which yields the normalization factor $C_{no\_absorp} = 0.926$ barn$\cdot$meV$^{-1}$ (solid line) and the extinction coefficient $\alpha = 0.143$. Similar fit of the absorption corrected DCS data [dashed line in Fig. S3D] yields the normalization factor $C_{absorp} = 0.42$ barn$\cdot$meV$^{-1}$ (solid line) and $\alpha = 0.083$. Fig. S3, E and H show the final normalized inelastic scattering cross-section in the (HHL) scattering plane summed within the energy range $E = [0.15, 1.5]$ meV, normalized without and with the absorption correction, respectively. The total scattering intensity averaged over the first Brillouin zone, within $Q_{HH} = [0, 1]$ and $Q_{00L} = [0, 2]$, for data without and with the absorption corrections is plotted in Fig. S3G and Fig. S3H, respectively. As we can see, the final integrated intensity is $6.5\mu_B^2 \cdot \text{meV}^{-1}/\text{Yb}$ for data without the absorption correction, and $7.5\mu_B^2 \cdot \text{meV}^{-1}/\text{Yb}$ for data with the absorption correction, with the mean square deviation of only $\approx 10\%$. This is because the diffraction peaks, which we use for normalization, are subject to the absorption correction on par with the inelastic spectrum that we normalize, so that accounting for absorption changes both the measured inelastic intensities and the value of the normalization factor, $C$. Hence, the absolute values of the normalized intensity remain nearly unaffected by the absorption correction, as they are only sensitive to the difference in absorption between the elastic diffraction peaks used for normalization and the inelastic scattering intensities, and not to the absolute impact of absorption, which is much larger than this difference. This is an advantage of the approach to absolute normalization that we have employed.

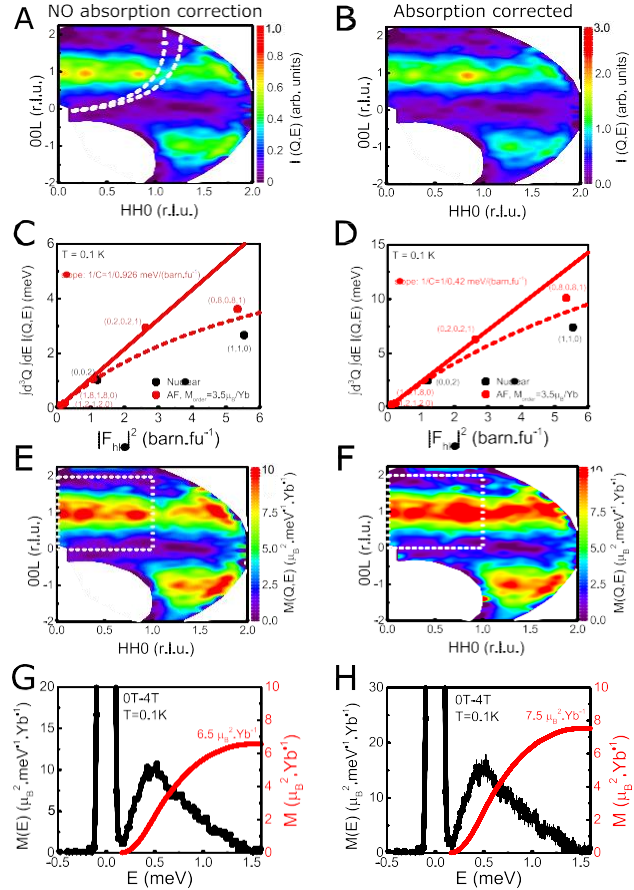

FIG. S3. **Effect of absorption in Yb$_2$Pt$_2$Pb**. (A), (B) Inelastic neutron intensity in the (HHL) scattering plane integrated within the energy range $E = [0.15, 1.5]$ meV measured at zero field, with the 4 T high field data subtracted, without the absorption correction (A) and with the absorption correction (B). The white dashed lines in (A) indicate the region where the measured intensity has largest suppression due to absorption, which is corrected in (B). The overall intensity is also enhanced upon accounting for absorption in (B). (C), (D) Comparison of the calculated nuclear and magnetic structure factors and the experimental integrated diffraction peak intensities at 0.1 K measured on DCS without (C) and with (D) the absorption correction. The solid and dashed lines are as in Fig. S2, and are explained in the text. (E), (F) Normalized inelastic scattering in the (HHL) scattering plane integrated within the energy range $E = [0.15, 1.5]$ meV measured in 0 T, with the 4 T high field data subtracted, without the absorption correction (E) and with the absorption correction (F). (G), (H) Energy dependence of the scattering averaged over $Q_{HH} = [0, 1]$, and $Q_{00L} = [0, 2]$, as indicated by the dashed rectangles in (E) and (F). The red lines in (C) and (D) are the energy integrated fluctuating moments without the absorption correction (G) and with the absorption correction (H).

### G. Sum rule

Components of the total fluctuating Yb moment, $M_\alpha$, are related to those of the effective spin-1/2, $S^\alpha$, via the effective $g-$factor,

$$M_\alpha = g_\alpha^{eff} \mu_B S^\alpha, \text{ where } (S^\alpha)^2 = 1/4, \tag{S5}$$

$\alpha = x, y, z$. The zero moment sum rule for the dynamical spin correlation function, $S^{\alpha\alpha}(\boldsymbol{Q}, E)$, in general (for $S \geq 1$) governs the sum of contributions from all three spin polarizations,

$$\sum_\alpha \int \frac{d^3\boldsymbol{Q}}{V^*} \int dE S^{\alpha\alpha}(\boldsymbol{Q}, E) = S(S+1), \tag{S6}$$

where integrations are over the first Brilluoin zone of volume $V^*$ and over all energies [40]. However, for $S = 1/2$, $(S^\alpha)^2 = 1/4$ is a $c-$number, and the sum rule is a combination of three independent sum rules,

$$\int \frac{d^3\boldsymbol{Q}}{V^*} \int dE S^{\alpha\alpha}(\boldsymbol{Q}, E) = (S^\alpha)^2 = 1/4, \ \alpha = x, y, z, \tag{S7}$$

one for each polarization. This observation allows us to relate the absolute magnetic scattering intensity measured in Yb$_2$Pt$_2$Pb with that calculated for the effective spin-1/2 XXZ model Hamiltonian.

Since we have been able to express the scattering intensity in absolute units of magnetic scattering cross-section, as shown in Fig. 4A in the manuscript, it is possible to demonstrate the evolution with temperature of the ordered moment $M_{\text{Ordered}}^2$ (black circles), fluctuating moment $M_{\text{Fluctuating}}^2$ (red circles) and the total moment $M_{\text{Total}}^2$ (blue diamonds). In the paramagnetic state at T$\geq$T$_N$, there is no static magnetic order, and all magnetic moments are fluctuating. The fluctuating moments are determined by the total inelastic scattering averaged over the first Brillouin zone and integrated in energy, as shown for different temperatures in Fig. 4A of the manuscript. With the onset of magnetic order at T$_N$=2.07 K, magnetic diffraction peaks appear and their intensity increases with the decreasing temperature, while at the same time there is a corresponding drop in the magnitude of the fluctuating moments, although there is a considerable fluctuating moment even at 0.1 K. The total moment, $M_{\text{Total}}^2 = M_{\text{Ordered}}^2 + M_{\text{Fluctuating}}^2$, is at most weakly temperature dependent, and within the error bar of our neutron measurements is constant and consistent in magnitude with the value expected for the ground state Yb doublet, $M_{1-10}^2 = 3.95^2\mu_{\text{B}}^2/\text{Yb}$, which is indicated by the thick gray line in Fig. 4A in the manuscript.

## II. CRYSTAL ELECTRIC FIELD SPLITTING OF THE YB $J = 7/2$ MULTIPLET AND THE SINGLE ION ANISOTROPY

### A. Crystal structure, multiplet splitting, and two Yb sublattices

Yb single ion anisotropy is very important for understanding the magnetic properties and the one dimensional spin dynamics of $Yb_2Pt_2Pb$. The electronic configuration of the Yb atom is $[\text{Xe}]4f^{14}6s^2$, and therefore $Yb^{3+}$ ions in $Yb_2Pt_2Pb$ have a single hole in the nearly filled $4f$ ($L = 3$) shell, which renders it a net total spin $S = 1/2$. The strong spin-orbit interaction, one of the strongest in the lantanide series, $\Lambda \sim 380$ meV, mandates that the total angular momentum, $\boldsymbol{J = L + S}$, rather than $\boldsymbol{S}$, or $\boldsymbol{L}$, is a good quantum number, and that the total angular momentum of the ground state multiplet is $J = 7/2$, with the first excited multiplet, $J = 5/2$, about 1.2 eV higher in energy [44]. In the $Yb_2Pt_2Pb$ layered crystal structure with 2 formula units and, correspondingly, two $a - b$ layers per unit cell, the Yb atoms occupy two inequivalent sites, labeled Yb1 (red) and Yb2 (blue) in Fig. S4A and Fig. S4B, which stagger along the $c-$axis. Both of the Yb sites have the same $C2v$ (m2m) point group symmetry, where each Yb ion is surrounded by six nearest neighbor Pt atoms (three above and three below the Yb-Yb plane), and four nearest neighbor Pb atoms (two above and two below the Yb-Yb plane) [5]. These Pt and Pb atoms make an anisotropic local environment of static charges with markedly uniaxial anisotropy, which determines the magnetic ground state of each of the Yb ions.

The eightfold degeneracy of the $J = 7/2$ multiplet of $Yb^{3+}$ on both Yb1 and Yb2 sites is lifted by the crystal electric field (CEF), which in this point group symmetry results in a manifold of four doublet states. We have calculated the resulting CEF splitting of the $J-$multiplet based on the point charge model, using the McPhase software [45–48]. Since $Yb_2Pt_2Pb$ is a good metal, it is hard to assign an accurate point charge to the ions, and so we have simply used charges corresponding to the average valences for Pt and Pb, of $Pt^{2+}$ and $Pb^{4+}$. This assumes that the collectivized electrons are in a good Bloch wave states and are evenly spread throughout the crystal, providing a homogeneous and isotropic negative charge background that does not contribute to the anisotropic crystal electric field (the screening length is larger that the inter-atomic

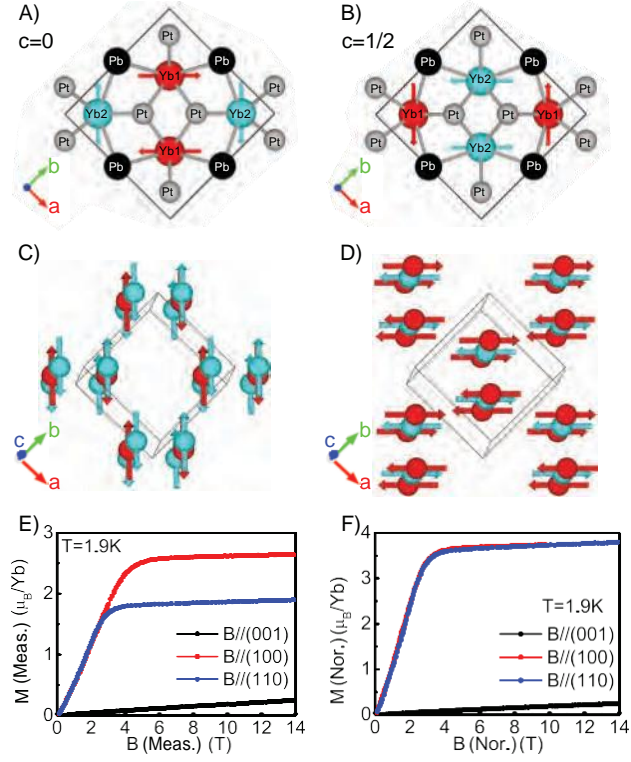

FIG. S4. **Yb crystal environment and magnetic moments in Yb$_2$Pt$_2$Pb.** (A), (B) Chemical environment for Yb1 (red) and Yb2 (blue) sites in the $c = 0$ (A), and $c = 1/2$ (B) plane, respectively. The nearest Pt (gray) and Pb (black) neighbors are located below and above the corresponding $c = 0$ and $c = 1/2$ plane [5]. Red and blue arrows indicate magnetic moment directions consistent with the point charge model calculation [45–48]. (C), (D) Simplified schematics of the two magnetic sublattices, with Yb moments along (-110) direction [sublattice 1, (C)], and along (110) direction [sublattice 2, (D)], respectively. (E) Experimentally measured magnetization at T = 1.9 K, normalized to the Yb$_2$Pt$_2$Pb formula unit, for magnetic field B applied along the three principal crystal directions, (110) [blue], (100) [red], and (001) [black]. (F) Magnetization normalized with the account of the two sublattice model illustrated in (A)-(D), as explained in the text. Plotted in blue is $M_{110}$(Nor.) vs. $B_{110}$(Nor.) calculated with Eq. (S11a), in red $M_{110}$(Nor.) vs. $B_{110}$(Nor.) calculated with Eq. (S11b), and in black $M_{001}$(Nor.) vs. $B_{001}$(Nor.) calculated with Eq. (S11c). The saturation moment for field along the diagonal, (110) direction, is $M_{110}$(Nor.) $\approx 3.8\mu_B$/Yb, which is much larger than the moment for B = 14T along the crystal (001) direction, $M_{001}$(Nor.) $\approx 0.25\mu_B$/Yb, at 1.9 K. The anisotropy of the magnetic susceptibility at 1.9 K, obtained from the ratio of the two magnetization slopes in the linear region, B $\lesssim$ 2T, indicates the effective g-factor anisotropy, $\frac{g_\parallel^{eff}}{g_\perp^{eff}} = 7.5(4)$, via $\frac{\chi_{110}(Nor)}{\chi_{001}(Nor)} = \left(\frac{g_\parallel^{eff}}{g_\perp^{eff}}\right)^2$.

distances), not an unreasonable approximation for a good metal.

Calculations show that both Yb1 and Yb2 sites have energetically well separated ground state doublets that are almost pure eigenstates of $|m_J = \pm 7/2\rangle$. As shown in Fig. S4A and Fig. S4B, the high symmetry directions for the Yb1 or Yb2 sites do not coincide with the crystalline $a$, $b$, or $c$ axes. For the Yb1 sites in the $c = 0$ plane [red Yb ions in Fig. S4A], the CEF states most nearly diagonalize if the local $\boldsymbol{J}$-quantization $z$-axis is chosen along the $(1, -1, 0)$ direction [Fig. S4C], and the resulting CEF levels are,

$$|E_{\pm}\rangle_0 = 0.992|\pm 7/2\rangle + 0.100|\pm 3/2\rangle + 0.064|\mp 1/2\rangle + 0.032|\mp 5/2\rangle, \tag{S8a}$$

$$|E_{\pm}\rangle_1 = -0.082|\pm 7/2\rangle + 0.300|\pm 3/2\rangle + 0.355|\mp 1/2\rangle + 0.881|\mp 5/2\rangle, \tag{S8b}$$

$$|E_{\pm}\rangle_2 = 0.092|\pm 7/2\rangle + 0.785|\pm 3/2\rangle + 0.420|\mp 1/2\rangle - 0.446|\mp 5/2\rangle, \tag{S8c}$$

$$|E_{\pm}\rangle_3 = -0.153|\pm 7/2\rangle + 0.832|\pm 3/2\rangle - 0.533|\mp 1/2\rangle + 0.005|\mp 5/2\rangle, \tag{S8d}$$

with the excited levels, $|E_{\pm}\rangle_1$, $|E_{\pm}\rangle_2$ and $|E_{\pm}\rangle_3$, separated from the ground doublet, $|E_{\pm}\rangle_0$, by energies $\Delta_1 = 37$ meV, $\Delta_2 = 64$ meV, and $\Delta_3 = 94$ meV, respectively. This corresponds to the crystal-field Hamiltonian,

$$H = \sum_{l,|m| \leq 2l} B_{2l}^m O_{2l}^m(\boldsymbol{J}), \tag{S9}$$

where $O_{2l}^m(\boldsymbol{J})$ are the Stevens operators [46], and the non-zero crystal-field parameters are, $B_2^0 = -2.242230$, $B_2^2 = -1.890630$, $B_4^0 = -0.002223$, $B_4^2 = 0.021004$, $B_4^4 = -0.046010$, $B_6^0 = 0.000008$, $B_6^2 = -0.000015$, $B_6^4 = 0.000124$, and $B_6^6 = -0.000150$.

Consequently, the magnetic properties of Yb$_2$Pt$_2$Pb at low temperatures must be dominated by the Yb$^{3+}$ ground state doublet, $|E_{\pm}\rangle_0$. Based on the CEF levels of Eqs. (S8), the calculated Yb magnetic moments for the Yb1 $(1 - 10)$ sublattice in $B = 14$T magnetic field applied along different directions are,

$$M_{\bar{1}10} \simeq \pm 3.95 \mu_B/\text{Yb}, \tag{S10a}$$

$$M_{110} \simeq \pm 0.2 \mu_B/\text{Yb}, \tag{S10b}$$

$$M_{001} \simeq \pm 0.3 \mu_B/\text{Yb}. \tag{S10c}$$

The moment along any direction other than $(1 - 10)$ is very small, indicating strong Ising-like magnetic anisotropy. Although the point charge model for Yb$_2$Pt$_2$Pb may be expected to be not as quantitative as it would be for an insulating magnetic material [49] due to the screening effect of the conduction electrons, the calculated magnetic moments are in a very good quantitative agreement with the M(B) anisotropy found in magnetization measurements, as we discuss below [Fig. S4, C and D]. Moreover, the calculated CEF splitting of the ground state doublet [Eq. (S8)], $\Delta_1$, is in good agreement with the crystal field excitation observed in our supplementary neutron scattering measurements.

Since both Yb sites share the same point symmetry and have very similar Yb-Pt and Yb-Pb distances, the CEF levels and the multiplet wave functions, including the ground state doublet, differ very little for the Yb2 sites from those calculated for Yb1 sites, Eqs. (S8). However, the crystal chemical environment of the Yb2 site is rotated by 90° with respect to that of the Yb1 site in the same ($c = 0$, or $c = 1/2$) plane. Therefore, the local $\boldsymbol{J}$-quantization axes and the magnetic moment directions calculated for the Yb2 sites in the $c = 0$ plane are now along the $(110)$ direction, as indicated by the blue arrows in Fig. S4A. Similarly, for the $c = 1/2$ plane, the moments are along $(110)$ for the Yb1 sites, and $(1-10)$ for the Yb2 sites [Fig. S4B]. Hence, there are two sets of Yb moments in Yb$_2$Pt$_2$Pb, forming two sublattices whose moments are orthogonal, directed along $(1-10)$ for sublattice 1 [Fig. S4C], and along $(110)$ for sublattice 2 [Fig. S4D].

The existence of these two orthogonal magnetic sublattices provides consistent and very natural explanation of the anisotropic magnetization in Yb$_2$Pt$_2$Pb, as has been proposed in [8, 50, 51]. The measured magnetization with magnetic field along the $(100)$, $(110)$ and $(001)$ principal axes [5, 6] are plotted in Fig. S4E, where the normalization to the units of $\mu_B/$Yb was obtained based on the total number of Yb atoms in the sample. However, the large CEF splitting ($\Delta_1 \approx 37$ meV), which isolates the ground state doublet, means that the magnetic single ion Ising anisotropy is much stronger than the maximum field, B = 14 T, of the DC magnetization measurements. Hence, the magnetization is weakly sensitive to magnetic field component perpendicular to the local $\boldsymbol{J}$-quantization axis, and only half of the Yb moments could be fully polarized when magnetic field is applied along the $(110)$ direction. Thus, for B $\parallel (110)$, the measured saturation magnetization, $M_{110}$(Meas.) $= 1.9\mu_B/$Yb, mainly originates from the sublattice 2 with moments along $(110)$, while the other sublattice, with moments along the $(1-10)$ direction, contributes very little. As only half of the Yb moments are contributing, $M_{110}$(Meas.) must be corrected by the factor of two [8, 50, 51], leading to the saturation moment $M_{110}$(Nor.) $= 3.8\mu_B/$Yb, close to the value calculated by McPhase, Eq. (S10a). When magnetic field is applied along the $(100)$ direction, $45°$ away from the Ising easy-axis direction of either of the two magnetic sublattices, all Yb magnetic moments are polarized along their own Ising axes, but with only a projection of the magnetic field on these axes being effective, so the measured saturation field along the $(100)$ direction, B$_{100}$(Meas.), is $\sqrt{2}$ times larger than the saturation field when magnetization is measured along the $(110)$ direction, B$_{110}$(Meas.) [8, 50, 51]. Similarly, the measured saturation moment, $M_{100}$(Meas.) $= 2.7\mu_B/$Yb, is the $45°$ projection of both $M_{110}$(Nor.) and $M_{1\bar{1}0}$(Nor.), which are in fact $\sqrt{2}$ times larger [8, 50, 51]. Magnetic field along the crystal $(001)$ direction is perpendicular to all Ising axes, and no moments in either sublattice are substantially polarized in fields as low as 14 T. Consequently, the measured magnetization, $M_{001}$(Meas.) $= 0.25\ \mu_B/$Yb, at 14 T is very small. We thus establish the following relationships between the actual magnetization of Yb ions, which we call normalized magnetization $M$(Nor.), and the effective magnetic field acting along the Ising axis, B(Nor.), and the experimentally measured magnetization $M$(Meas.) and the applied magnetic field B(Meas.),

$$M_{110 \text{ or } \bar{1}10}(\text{Nor.}) = M_{110}(\text{Meas.}) \times 2 \, [= 3.8 \mu_B/\text{Yb@14T}], \quad B_{110}(\text{Nor.}) = B_{110}(\text{Meas.}), \quad \text{(S11a)}$$

$$M_{110 \text{ or } \bar{1}10}(\text{Nor.}) = M_{100}(\text{Meas.}) \times \sqrt{2} \, [= 3.8 \mu_B/\text{Yb@14T}], \quad B_{110}(\text{Nor.}) = B_{100}(\text{Meas.})/\sqrt{2}, \quad \text{(S11b)}$$

$$M_{001}(\text{Nor.}) = M_{001}(\text{Meas.}) \, [= 0.25 \mu_B/\text{Yb@14T}], \quad B_{001}(\text{Nor.}) = B_{001}(\text{Meas.}). \quad \text{(S11c)}$$

Fig. S4F shows the normalized magnetization against the accordingly normalized field, as defined by Eqs. (S11). We should note that because of the 4-fold macroscopic symmetry of the two Yb sites and the corresponding two Ising axes, there is always a small contribution coming from the Yb magnetization induced by the field component perpendicular to its Ising axis when magnetic field is applied in the *ab* plane. For example, $M_{110}$(Meas.) contains small transverse contribution from the $(1 - 10)$ sublattice, and $M_{100}$(Meas.) has similar transverse contributions from both sublattices, which are neglected in Eqs. (S11a)-(S11b). However, based on the calculation [Eqs. (S10a)-(S10c)] this contribution is even smaller than the magnetization $M_{001}$(Nor.) for B along the (001) direction, which justifies our neglect of this component.

It is also impossible to conclude from the DC magnetization measurements alone, which of the two sublattices (i.e. which of the two Yb sites, Yb1, or Yb2, in a given *ab*-layer) has moments along the (110), and which along the $(1 - 10)$ direction. Given the two-sublattice structure, magnetization cannot distinguish whether moments are parallel or perpendicular to the putative Yb-dimer bond direction [5, 8]. Our point charge model CEF calculations indicate, however, that the Yb Ising moments are perpendicular to the nearest neighbor Yb-Yb bond [Fig. S4], which is consistent with the zero field magnetic structure determined from powder neutron diffraction [18]. We note that the magnetic structure that we infer in Yb$_2$Pt$_2$Pb is quite different from the magnetic structure of U$_2$Pt$_2$Sn, and of the SSL system GdB$_4$ [52, 53].

Finally, two-sublattice structure helps to understand the one-dimensionality of Yb$_2$Pt$_2$Pb. The nearest neighbor Yb-Yb pairs in the crystalline *ab*−plane belong to different sublattices [Fig. S4A–D]. That the two sublattices are easily decoupled even by small magnetic fields and are magnetized separately, indicates that the inter-sublattice interactions are very weak. On the other hand, the distance between the nearest neighbor Yb pairs within each sublattice in the *ab*−plane is $a = 7.76$Å [Fig. S4A–D]. Therefore, these Yb are only weakly coupled. In contrast, the distance between Yb from the same sublattice along the crystal *c*−axis is only $\simeq c/2$, $\approx 3.5$Å [Fig. S4A–D], and therefore their coupling in this direction is stronger. This immediately leads to one dimensional Yb chains, or ladders, along the crystallographic *c*−axis, and thus the one dimensional dispersion with respect to wave vector $Q_{00L}$, as seen in Fig. S8, C and D.

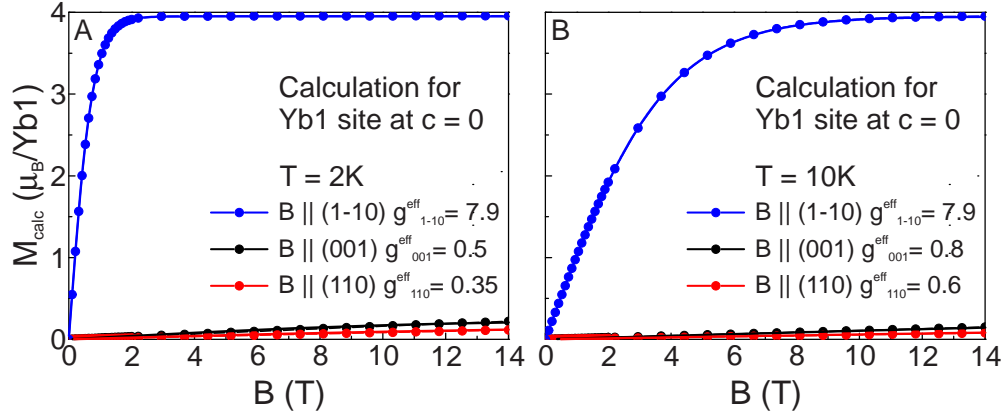

FIG. S5. **Calculated single-ion magnetization based on the CEF model in Yb$_2$Pt$_2$Pb.** Magnetization for an isolated Yb ion belonging to the Yb1 (1-10) sublattice as a function of magnetic field B up to 14 T for the three directions of the field at T = 2 K (A) and at T = 10K (B). Solid curves show fits to the Brillouin function for S=1/2, which determine the effective spin-1/2 $g-$ factors describing the ground state doublet.

## B. Magnetization and single ion g-factors

The point charge CEF model also allows us to determine the single ion effective spin-1/2 $g-$factors for the Yb ground state doublet. To this end, we calculate magnetization curves, M(B), of the Yb CEF multiplet [Eq. S8] for some finite temperature T for magnetic field along the three principal crystal axes, which we then fit to a Brillouin function with $S = 1/2$,

$$M_\alpha = g_\alpha^{eff} \mu_B S B_S(x), \ \ B_S(x) = \frac{2S+1}{2S} \coth\left(\frac{2S+1}{2S}x\right) - \frac{1}{2S}\coth\left(\frac{1}{2S}x\right), \ \ x = \frac{g_\alpha^{eff} \mu_B S B}{k_B \text{T}}. \quad (S12)$$

Here $\alpha$ denotes the crystallographic direction, $M_\alpha$ is the magnetization per Yb, $\mu_B$ is Bohr magneton, $k_B$ is Boltzman constant, and $g_\alpha^{eff}$ is the effective $g-$factor for magnetic field B in the direction $\alpha$, which is obtained from the fit. Calculations for the Yb1 site with the Ising axis along the $(1-10)$ direction for T = 2 K and T = 10 K (symbols) are shown together with the corresponding fits (lines) in Fig. S5. The obtained $g-$factor values,

$$g_\parallel^{eff} = g_{1-10}^{eff} = 7.9 \quad (S13a)$$

$$g_{\perp,ab}^{eff} = g_{110}^{eff} \lesssim 0.6 \quad (S13b)$$

$$g_{\perp,c}^{eff} = g_{001}^{eff} \lesssim 0.8, \quad (S13c)$$

are in good agreement with the experimental results that we report, and in particular with nearly completely longitudinal character of the magnetic excitation spectrum, which is proportional to $\left(g_\alpha^{eff}\right)^2$.

The magnetization curves measured in Yb$_2$Pt$_2$Pb for different directions of magnetic field at T = 1.9 K shown in Fig. S4, E and F, allow us to make an experimental estimate of the effective g-factor anisotropy from the anisotropy of magnetic susceptibility at 1.9 K. Using a simple Curie law for a paramagnet, $\frac{\chi_{110}}{\chi_{001}} = \frac{\left(g_{110}^{eff}\right)^2}{\left(g_{001}^{eff}\right)^2}$, we obtain $\frac{g_{\parallel}^{eff}}{g_{\perp}^{eff}} = 7.5(4)$, which agrees remarkably well with the calculations, Eq. (S13), even though susceptibility at 1.9 K might be markedly affected by the antiferromagnetic correlations that are neglected in the single ion calculation (S13). The magnetization curves for B$\parallel(1-10)$ at T = 0.09 K and 1.5 K are shown in Fig. S6A, while in Fig. S6B the measured magnetization is compared with the theoretical magnetization curves for the spin-1/2 XXZ model Eq. S18, obtained using the quantum transfer matrix approach suitable for finite temperature calculations, similar to the T-dependent magnetic susceptibility shown in Fig. 4B of the manuscript. Both the temperature dependence of magnetic susceptibility (it has also been reported in [6]), and the magnetization curves agree quite well with the calculation for the effective nearest neighbor spin-1/2 XXZ Hamiltonian Eq. S18 with $g^{eff} = 7$, $\Delta = 2.6$ and J = 0.205 meV. The largest discrepancy in M(H) occurs near the saturation field, where the high energy four spinon contribution that is absent in the approximate nearest neighbor model becomes important (cf Fig. 1 and the discussion in the manuscript).

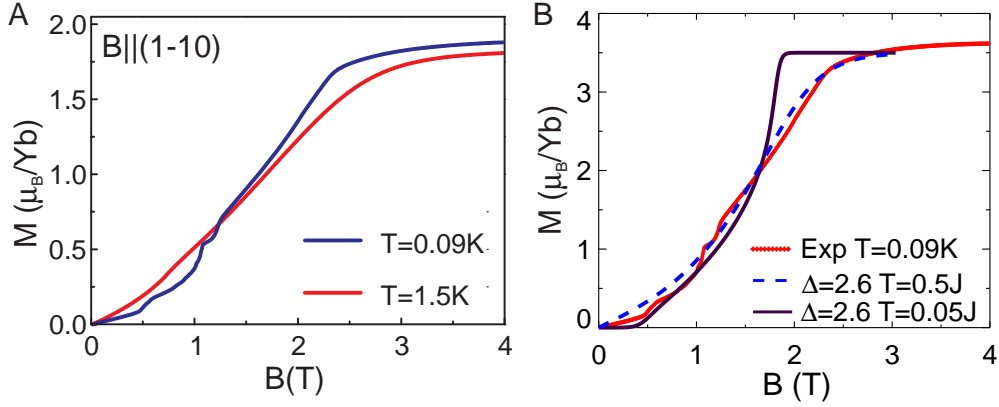

FIG. S6. (A) Magnetization measured on a typical single crystal of Yb$_2$Pt$_2$Pb in magnetic field B along the $(1-10)$ lattice direction at T = 0.09(1)K (dark blue) and 1.5(1) K (red), without the two-sublattice correction (S11). (B) Magnetization calculated for the effective spin-1/2 XXZ Hamiltonian (S18) with $g^{eff} = 7$, $\Delta = 2.6$ and J = 0.205 meV for T = 0.05J $\approx$ 0.1 K (dark solid line) and T = 0.5J $\approx$ 1.2 K (blue broken line) using the quantum transfer matrix approach suitable for finite temperature calculations, as described in Methods. Red symbols show magnetization measured in Yb$_2$Pt$_2$Pb at T$\approx$ 0.09 K, same as in panel A, but corrected by a factor 2 according to Eq. (S11a) and by subtracting a small linear component accounting for the contribution of the perpendicular sublattice.

## C. Crystal field excitation

In order to verify directly the hierarchy of the Yb crystal field levels predicted by the point charge model, we have carried out additional measurements on CNCS, using the incident neutron energy of 37 meV, Fig. S7. Inelastic neutron spectra collected at T = 175 K reveal a non-dispersing excitation with an energy $E_c = 25.2(1)$ meV [Fig. S7], which is in reasonably good agreement with the splitting, $\Delta_1 = 37$ meV, between the ground state and the first excited doublet predicted by the point charge model. Unlike the phonon intensity that increases with the increasing magnitude of the wave vector transfer, $Q$, the intensity of the observed excitation decreases at large $L$ (large $Q$), which confirms its magnetic origin. This result renders further direct experimental credibility to our CEF calculations, which are already strongly supported by the magnetization and the magnetic susceptibility data. We have to note that in the early reports by some of the present authors [5], a broad contribution to the low-temperature specific heat was found, whose Schottky-peak model analysis indicated the lowest doublet splitting of $\Delta_1 \approx 7$ meV, significantly lower than the excitation observed here by neutron scattering. However, a more recent, closer inspection of that data lead us to conclude that the "Shottky" peak (e.g. in Fig. 6 of Ref. [5]) likely results from a systematic error in subtracting the Debye contribution [18]. Hence, the heat capacity results can only be understood to indicate a lower limit, $\Delta_1 \gtrsim 7$ meV, on the CEF doublet splitting. Moreover, the temperature dependence of the inverse magnetic susceptibility, which was measured in Ref. [5] up to 800 K, indicates a change of slope at around 300 K, which is fully consistent with $E_c = 25.2(1)$ meV observed here.

Although in principle it is possible that the Yb1 and Yb2 sites could have slightly different crystal field splitting, we do not resolve any such additional peak structure, or any considerable thermal broadening of this level at 175 K, within the $\approx 3.6$ meV energy resolution of this high energy CNCS measurement, Fig. S7B. Taken together, our experimental data confirm that the first excited CEF level in Yb$_2$Pt$_2$Pb is energetically well separated from the ground state doublet, and provide strong support for our CEF point charge model calculations. Thermal population of the excited doublets is negligible within the temperature range, $T \lesssim 100K$, of the measurements that we report in the manuscript.

More precisely, the lowest splitting of the CEF doublet manifold, $\Delta_1 \sim 30$ meV, implies that there is no significant thermal occupation of the first excited CEF level for temperatures T lower than $\approx 300$ K, or magnetic fields less than $B \approx 300$ T. The data we present in both the main text of this paper and the supplement were measured well within this field ($B \leq 14T$) and temperature ($T \leq 100K$) range. Thus, the Ising anisotropy of the doublet ground state is robust, and protected by the magnitude of $\Delta_1$ over our full experimental field and temperature window. Neglecting higher energy states, we can separate the two orthogonally aligned Yb moments into two magnetic sublattices based on the direction of their Ising anisotropy axes. Shown in Fig. S4C is the magnetic sublattice with all the Yb1 sites in the $c = 0$ plane, and all the Yb2 sites in the $c = 1/2$ plane, with the magnetic moments pointing along the (-1, 1, 0) direction. All the Yb2 sites in the $c = 0$ plane and all the Yb1 sites in the $c = 1/2$ plane form a different, but equivalent magnetic sublattice, with all the magnetic moments pointing along the (1, 1, 0) direction [Fig. S4D]. It is worthwhile to notice that the originally proposed Shastry-Sutherland lattice (SSL) based on the crystal structure in the *ab* plane [5, 6, 8, 50, 51] is now broken, and instead each sublattice consists of square lattices with well separated ladders, or pairs of chains, along the crystal *c*−axis. Thus it is not surprising that one dimensional dynamics with spinon excitations are observed along $Q_{00L}$ as shown in the manuscript.

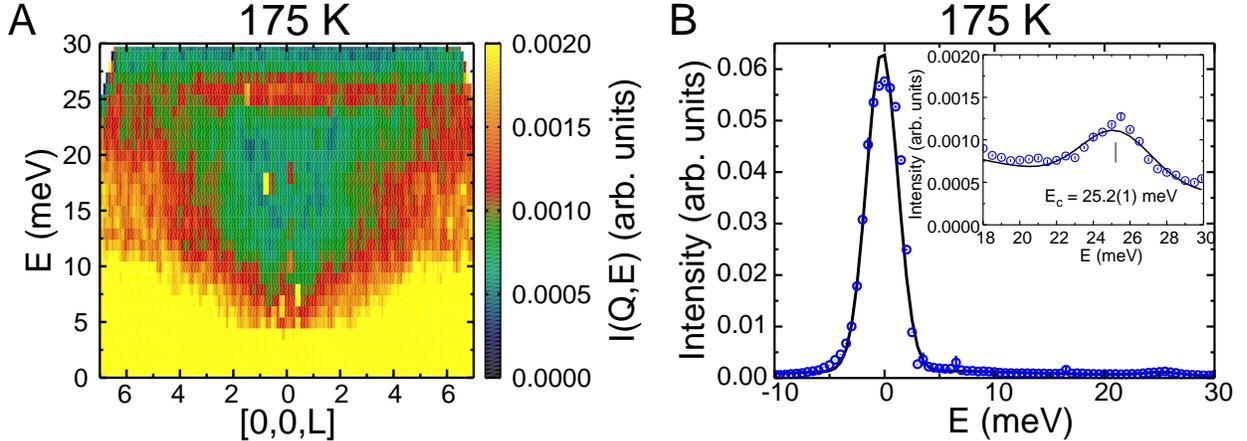

FIG. S7. **Crystal field excitation in Yb$_2$Pt$_2$Pb.** (A) Contour map of the inelastic scattering intensity measured at T $\approx$ 175K, with energy dispersion along the wave vector $Q_{00L}$, averaged over the perpendicular wave vector components within (HH) = [−5,5], (H−H) = [−0.15,0.15]. The scattering at high wave vectors and energies less than 20 meV is dominated by the aluminum sample holder and the cryogenic environment. (B) Energy dependent intensity averaged over the wave vector range (HH) = [−5,5], (H−H) = [−0.15,0.15], L = [−5,5]. The inset shows the enlarged plot of the CEF excitation around 25.2 meV.

## III. SUPPLEMENTARY THEORY

### A. The High-Energy Hamiltonian and the Orbital Exchange Interaction

The emergence of spinons in Yb$_2$Pt$_2$Pb becomes less counterintuitive when one steps back and considers the underlying high-energy *f*-electron Hamiltonian, which can be expressed in the form of a one-dimensional Hubbard model with strong spin-orbit coupling [27],

$$H = \sum_{<n,m,j,j'>} t_{nm}^{jj'} f_j^+(n) f_{j'}(m) + \sum_{<n,j>} U[f_j^+(n) f_j(n) - N]^2 + \sum_{<n,j>} \Delta_j f_j^+(n) f_j(n), \qquad (S14)$$

where $n, m$ number the Yb sites, the last term represents the crystal field, $j, j'$ are total angular momentum projections, and $N = 7$ is the occupation number of the 4*f*-shell with $J = 7/2$. The hierarchy of interactions is such that $U \gg t_{nm}^{jj'} \gtrsim \Delta_j$. Considering $t_{nm}^{jj'}$ and $\Delta_j$ as perturbations, we assume the axial symmetry of the unperturbed system so that the tunneling is diagonal in the total angular momentum projection. In the second order of perturbation in the tunneling matrix element we obtain the exchange interaction in the form,

$$H_{exchange} = \frac{t_{nm}^{jj} t_{nm}^{kk}}{U(2N+1)} f_j^+(n) f_k(n) f_k^+(m) f_j(m), \quad \sum_j f_j^+(n) f_j(n) = N. \tag{S15}$$

When all tunneling matrix elements $t^{jj}$ are equal, the exchange interaction is just a permutation operator, $P_{nm}$, acting on $(2J+1) \times (2J+1)$-dimensional space of sites $n$ and $m$,

$$H_{exchange} = \frac{t^2}{(2N+1)U} P_{nm}. \tag{S16}$$

This interaction has a high, $SU(2J+1)$ symmetry, and in this form it has been known for a long time in the theory of low dimensional magnetism [23–25], and is also frequently discussed in the theory of cold atoms (see, for example, [26]). The most striking feature of (S15) or its simplified form (S16) is that, unlike the Heisenberg-Dirac bilinear exchange $\sim J_n J_m$, it contains matrix elements between *all* angular momentum projections, in the latter case even with equal weight. Thus, it can be represented as a polynomial of order $(2J+1)$ in $J^+, J^-$ raising and lowering operators of the total angular momentum. The next step towards obtaining the effective low energy exchange Hamiltonian is to take into account the crystal field. If, as in our case, it is stronger than the exchange integral $\sim t^2/U$, then one can project (S15) on the lowest-energy multiplet (in our case it is the $j = \pm 7/2$ Kramers doublet) to obtain,

$$H_{eff} = \sum_{j,j'=\pm 7/2} \frac{t_{nm}^{jj} t_{nm}^{j'j'}}{U(2N+1)} f_j^+(n) f_{j'}(n) f_{j'}^+(m) f_j(m), \quad \sum_{j=\pm 7/2} f_j^+(n) f_j(n) = 1. \tag{S17}$$

The operators $S^\alpha = \sum f_j^+ \sigma_{jj'}^\alpha f_{j'}$ ($\sigma^\alpha$ being Pauli matrices, $\alpha = x, y, z$), subject to the occupation constraint, represent the components of a spin-1/2 operator. Hence, (S17) is equivalent to the S = 1/2 Heisenberg-Dirac XXZ Hamiltonian for pseudospins $S^\alpha$,

$$H = J \sum_n \left( S_n^x S_{n+1}^x + S_n^y S_{n+1}^y + \Delta S_n^z S_{n+1}^z \right), \tag{S18}$$

[Eq. (1) in the main text], replacing the $f$-electron Hamiltonian (S14) at low energies. The Hamiltonian (S17) accounts for the exchange of the two electrons between the $j = \pm 7/2$ states ($l = \pm 3$ states of the $f$-orbital) on the neighbor sites, Fig. 1D in the main text. For these states, strong spin-orbit coupling effectively quenches the electronic spin as an independent degree of freedom, simply dictating that it is co-aligned with the orbital angular momentum, $L$, which is the main contributor to the total angular momentum, $J$. Therefore, $U(2)$ pseudospin $S$ essentially describes the orbital degree of freedom [23] of the Yb $f$-electrons, and the effective spin-1/2 XXZ Hamiltonian (S18) represents the orbital-exchange Hamiltonian (S17). Although it has the familiar Heisenberg-Dirac-like bilinear form for the effective spin-1/2, in the general case it is anisotropic, and can also have non-diagonal elements which couple different spin components, $\sim S_n^\alpha S_m^\beta$. Most importantly, it contains matrix elements between the states that the bilinear exchange interaction of the form $\sim J_n J_m$ does not have.

### B. Spinon dispersion in spin-1/2 XXZ model.

In the Heisenberg-Ising case the anisotropy parameter in the XXZ Hamiltonian, Eq. (S18), is $\Delta > 1$, and the spinon dispersion, $\varepsilon_s(q)$, acquires a finite gap, $\Delta_s$. Using the conventional parameterization of the algebraic Bethe ansatz solution [12–14],

$$\varepsilon_s(q) = I\sqrt{1 - k^2\cos^2(qa)} = \sqrt{I^2\sin^2(qa) + \Delta_s^2\cos^2(qa)}, \tag{S19}$$

where $q \in [0, \pi]$ is the spinon's one-dimensional wave vector, $a$ is the lattice spacing in the chain, and the spinon gap, $\Delta_s = I\sqrt{1 - k^2} = Ik'$, and the dispersion bandwidth parameter,

$$I = J\frac{K(k)}{\pi}\sqrt{\Delta^2 - 1}, \tag{S20}$$

are both encoded by the integral equation,

$$\frac{K(k')}{K(k)} = \frac{K(k')}{K(\sqrt{1 - k'^2})} = -\frac{1}{\pi}\ln\left(\Delta - \sqrt{\Delta^2 - 1}\right). \tag{S21}$$

Here $K(k)$ is the complete elliptic integral of the first kind with elliptic modulus $k$, and $k'$ is the complementary elliptic modulus. For small $k'$, $K(k')$ can be expressed as a power series,

$$K(k') = \frac{\pi}{2}\left(1 + \left(\frac{1}{2}\right)^2 k'^2 + \left(\frac{3}{8}\right)^2 k'^4 + O(k'^6)\right), \tag{S22}$$

while for sufficiently large $\Delta$ the logarithm in Eq. (S21) can be Taylor-expanded,

$$\frac{K(k')}{K(k)} = \frac{K(k')}{K(\sqrt{1 - k'^2})} = \frac{1}{\pi}\left(\ln(2\Delta) - \frac{1}{4\Delta^2} + O(\frac{1}{\Delta^4})\right). \tag{S23}$$

Equivalently, the integral equations relating $I$ and $\Delta$ with $k, k'$ are [12–14],

$$\Delta = \cosh\left(\pi\frac{K(k')}{K(k)}\right), \ I = J\frac{K(k)}{\pi}\sinh\left(\pi\frac{K(k')}{K(k)}\right) = JK(k')\frac{\sqrt{\Delta^2 - 1}}{-\ln\left(\Delta - \sqrt{\Delta^2 - 1}\right)}. \tag{S24}$$

Using the experimentally determined values $I = 0.322$ meV and $\Delta_s = 0.095$ meV, we obtain $k' = \Delta_s/I = 0.295$, $k = \sqrt{1 - k'^2} = 0.955$, and $K(k') \approx 1.606$, $K(k) \approx 2.638$, $K(k')/K(k) \approx 0.609$. Substituting these values in Eq. (S24), we obtain $\Delta \approx 3.46$, and $J \approx 0.116$ meV, so that $I \approx 2.78J$ instead of $\frac{\pi}{2}J$ in the isotropic Heisenberg case where $k' = 0$.

### C. The boundaries of 2- and 4-spinon continua.

In the case of non-interacting quasi-particles, the energy of a multi-particle state is simply given by a sum of energies of the constituent particles, while the total wave vector is a sum of their wave vectors by the law of quasi-momentum conservation. Such multi-particle states then form a continuum, whose upper and lower boundaries are determined by the maximum and the minimum values of the total energy for a given total momentum and with respect to all possible relative momenta of the constituent particles. Thus, for the case of the 2-particle spinon continuum we have,

$$\varepsilon_{2s}(q) = min, max \{\varepsilon_s(q_1) + \varepsilon_s(q_2)\}, \tag{S25}$$

where the extremum is sought for the fixed total wave vector of the two-particle state, $q = q_1 + q_2$, of two spinons with wave vectors $q_1$ and $q_2$. The condition can be eliminated by solving the conservation law for the wave vector of one of the particles, leading to,

$$\varepsilon_{2s}(q) = min, max \{\varepsilon_s(q_1) + \varepsilon_s(q - q_1)\} = min, max \{\varepsilon_s(q/2 + q') + \varepsilon_s(q/2 - q')\}, \tag{S26}$$

where $q' = (q_1 - q_2)/2$ is half the spinons relative momentum, $q_{1,2} = q/2 \pm q'$. This results in the following minimum/maximum condition,

$$\varepsilon'_s(q/2 + q') = \varepsilon'_s(q/2 - q'), \text{ or, } \varepsilon'_s(q_1) = \varepsilon'_s(q_2), \tag{S27}$$

which essentially requires that spinons at the lower (upper) boundary have equal velocity, or that momentum of one of the spinons is at the boundary of the 1-spinon Brillouin zone (BZ), i.e. $q_1 = q/2 - q' = \pi$, and $q_2 = q/2 + q' = q - \pi \equiv q$, or vice versa. For the spinon dispersion given by Eq. (S19), the equal-velocity condition (S27) within the $[0, \pi]$ spinon BZ is only achieved for $q' = 0$, and therefore the upper and the lower boundaries of the 2-spinon continuum are defined by the following condition [13, 14, 16],

$$\varepsilon_{2s}^{low,up}(q) = min, max \{2\varepsilon_s(q/2), \Delta_s + \varepsilon_s(q)\}, \tag{S28}$$

where the minimum and the maximum values of the two expressions, $2\varepsilon_s(q/2)$ and $\Delta_s + \varepsilon_s(q)$, determine the lower and the upper boundary of the 2-spinon continuum, respectively. Near $q = \pi$ the lower boundary is given by $\Delta_s + \varepsilon_s(q)$, while near $q = 0$ by $2\varepsilon_s(q/2)$. The situation is opposite for the upper boundary. These boundaries are shown by solid white lines in Figure 2 of the manuscript.

The boundaries of the 4-spinon continuum are obtained using a similar extremum search procedure and applying the method of Lagrange multipliers, which leads to the same requirement of equal velocities. The upper boundaries, which in a range near $q = \pi/2$ are given by $max\{4\varepsilon_s(q/4), 4\varepsilon_s((q + 2\pi)/4)\}$, [14], are shown by broken white lines in Figure 2 of the manuscript.

## IV.  SUPPLEMENTARY NEUTRON SCATTERING DATA

### A.  Scattering in zero and high magnetic field and two magnetic sublattices

As discussed above, strong Ising anisotropy divides Yb atoms in $Yb_2Pt_2Pb$ into two orthogonal sublattices, with Ising axes along (110) and (1-10) directions, respectively. Magnetic field $B > B_c \approx 2.3$ T applied along (110), or (1-10) direction, can separately polarize moments in one of these two sublattices, so that only a single sublattice contributes to the measured inelastic neutron magnetic scattering at high fields. Here, we present the neutron scattering data at zero and 4 Tesla, which support this conclusion.

First, our neutron scattering measurements demonstrate that two magnetic sublattices with the orthogonal moments order independently below the Néel temperature $T_N = 2.07$ K, and that magnetic order of one of the two sublattices can be tuned separately by a magnetic field along the (1-10) direction. Shown in Fig. S8, A and B are the contour maps of the elastic neutron scattering intensity in the (HH, H-H, 1) scattering plane measured on CNCS at 1.5 K, in 0 T and 4 T field, respectively. This plane is perpendicular to the horizontal (HHL) scattering plane [Fig. S1, C and D], and the $Q_{H-H}$ dependence is only resolved by using the vertical position sensitivity of the CNCS detectors.  In zero field [Fig. S8A], two sets of satellite peaks, with $Q_1 = (\pm0.2, \pm0.2, 1)$ and $Q_2 = (\pm0.2, \mp0.2, 1)$, sit symmetrically around the (001) position. This indicates a $5 \times 5$ magnetic superstructure in the $ab$ plane in real space [18], and the propagation vectors $Q_1$ and $Q_2$ correspond to the respective AF ordering of the two sublattices.  Application of a 4 T magnetic field along the (1-10) direction polarizes the sublattice with magnetic moments along this direction, while the other, perpendicular sublattice appears essentially unaffected. As sublattice parallel to the field is magnetized, the diffraction intensity of the corresponding antiferromagnetic satellites at $Q_1 = (\pm0.2, \pm0.2, 1)$ decreases, and is transferred to the lattice Bragg positions by the arising ferromagnetic component. We indeed observe an increase of the (002) lattice Bragg reflection proportional to the square of magnetization. At 4 T, peaks at $Q_1$ are fully suppressed, while the other set of satellite peaks, at $Q_2$, remain nearly unchanged [Fig. S8B].

The two-sublattice behavior is further revealed by the inelastic neutron scattering data.  Shown in Fig. S8C is the contour map of the inelastic scattering intensity as a function of energy transfer and the wave vector $Q_{00L}$, averaged over $Q_{HH} = [0, 1]$ and $Q_{H-H} \approx [-0.1, 0.1]$, measured on DCS at 0.1 K in nominal magnetic field B = 0 T. The solid and dashed lines indicate the lower and upper boundaries of the two-spinon and four-spinon excitations, as explained in the main text.  The experiment was repeated with B = 4 T, as shown in Fig. S8D, and the energy and wave vector dependencies of the scattered intensity are very similar to those measured in zero field.  However, the absolute magnitude of the scattered intensity in 4 T is significantly lower than in zero field (the intensities in Fig. S8 C and D are normalized to units of $\mu_B^2 \cdot \text{meV}^{-1}$ per crystal unit cell with 8 Yb). Further quantitative comparison of the 0 T and the 4 T data is presented in Fig. S8E, which displays magnetic scattering intensity averaged over $Q_{00L} = [0, 2]$, $Q_{H-H} \approx [-0.1, 0.1]$, and $Q_{HH}$ near zero, $Q_{HH} = [-0.1, 0.1]$, as a function of energy transfer, $E$. A broad, asymmetric peak is observed in both 0 T and 4 T data at energies from about 0.15 meV to 1.5 meV. This is the spinon continuum, and it is considerably broader than the instrument resolution, which is about 0.1 meV, as estimated from the width of the elastic diffraction peaks. The shape of the broad peak at 4 T is similar to that at 0 T, but with about twice lower intensity. As we integrate this broad continuum with respect to energy up to an energy $E$ [Fig. S8F], we find that the integrated intensity at 0 T is almost exactly twice the integrated intensity at 4 T. The filled black symbols in Fig. S8F represent the 0 T integrated scattering intensity, the open red circles are twice the integrated scattering intensity at 4 T, and they overlap very closely with the 0 T data within the entire energy range covered, from 0.15 meV up to 1.75 meV.

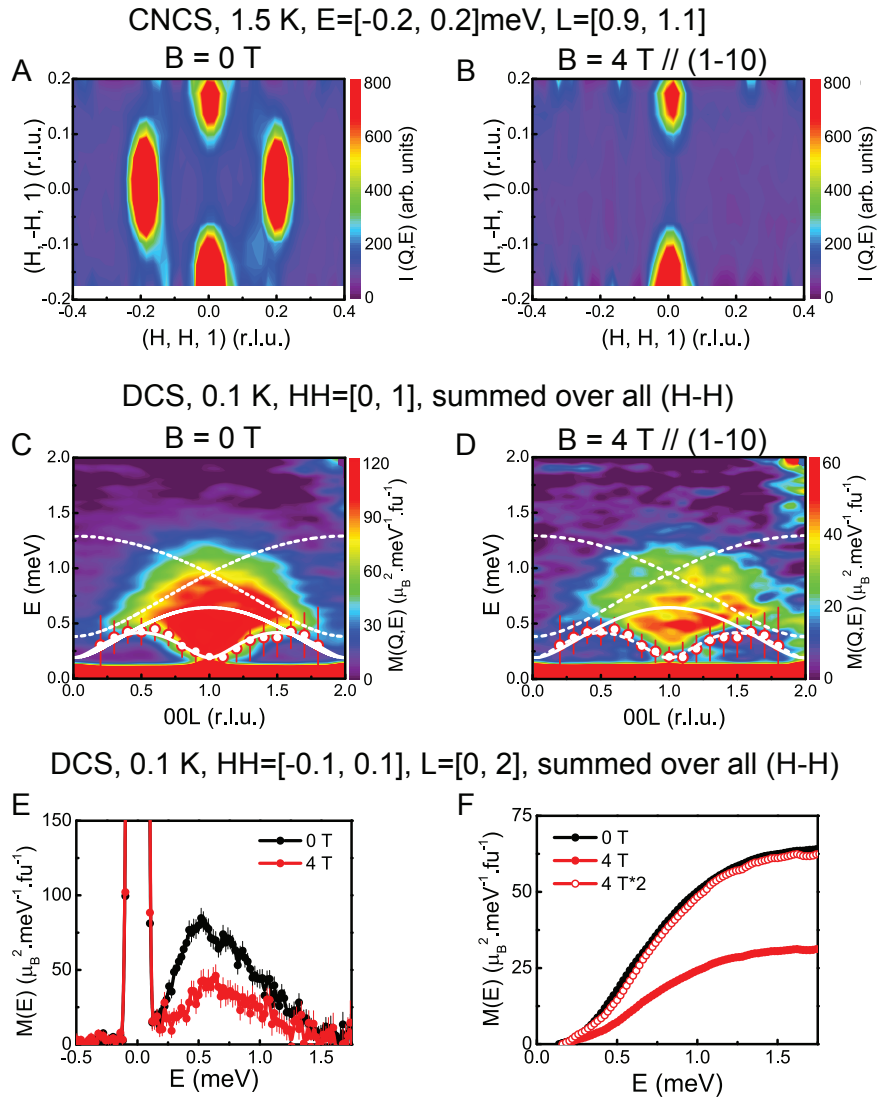

**FIG. S8. Magnetic neutron scattering from the two sublattices in Yb$_2$Pt$_2$Pb in 0 T and 4 T magnetic field.** (A), (B) Contour maps of the elastic diffraction in the (HH, H-H, 1) scattering plane within $E = [-0.2, 0.2]$ meV, and $Q_{00L} = [0.9, 1.1]$ measured on CNCS at 1.5 K in 0 T (A) and 4 T field along (1-10) (B). (C), (D) Contour maps of the inelastic scattering intensity as a function of energy transfer $E$ and wave vector $Q_{00L}$, within $Q_{HH} = [0, 1]$ and $Q_{H-H} \simeq [-0.1, 0.1]$, measured on DCS at 0.1 K in zero magnetic field (C), and in field B = 4 T along $(1-10)$ (D). The solid lines, dashed lines, and data points are explained in the main text. (E) Energy dependent scattering intensity in magnetic fields 0 T and 4 T, as indicated. The plot is obtained by averaging the neutron scattering intensity over $Q_{HH} = [-0.1, 0.1]$, and $Q_{00L} = [0, 2]$. (F) Energy integrated scattering intensity in 0 T and 4 T. The total scattering intensity in 0 T (black symbols) is nearly twice the intensity at 4 T (red symbols), as indicated by the open red circles.

Another remarkable observation is the notable absence of the coherent spin wave excitation of the fully polarized ferromagnetic sublattice at B = 4 T in Fig. S8D. Such spin wave is naturally expected in the fully saturated ferromagnetic state of a spin-1/2 Heisenberg, or XXZ chain at high field, and has been experimentally observed, e. g. in copper sulphate [15]. In $Yb_2Pt_2Pb$, however, such excitation, while theoretically present, and contributing to the system's energy and entropy in the effective spin-1/2 Hamiltonian, is not visible in experiment. The reason is, that it involves *transverse* fluctuations of the effective spins-1/2, and such fluctuations do not lead to any measurable magnetization fluctuations. The latter are almost entirely longitudinal by virtue of the strong anisotropy of the effective $g-$factor. Excitation of a transverse spin wave corresponds to lowering (raising) of the angular momentum of the Yb magnetic moment by $\Delta m_J = \pm 1$ via magnetic dipole interaction with a neutron, or a photon. However, such processes are ruled out by the nearly pure $|m_J = \pm 7/2\rangle$ nature of the Yb ground state doublet, which is protected by a large, $\gtrsim 25$ meV gap to the nearest excited state. Only orbital-exchange processes with $\Delta m_J = 0$ are allowed, as illustrated in Fig. 1 of the main text. This invisible transverse spin wave excitation of the ferromagnetically polarized sublattice at high-field, a "dark magnon" quasi-particle, perhaps offers a condensed matter analogue of the weakly interacting massive particles (WIMPs) of the cosmological dark matter.

The above observations indicate that, while in the absence of magnetic field both sublattices contribute equally to magnetic scattering, when magnetic field is applied along one of the orthogonal Ising directions [here $\mathbf{B}\|(1-10)$] and polarizes magnetic moments in the corresponding sublattice, only fluctuations of the other sublattice, with moments perpendicular to the field direction and essentially unaffected by the magnetic field, contribute to the magnetic scattering intensity. Hence, the difference between the 0 T and the 4 T data yields the net contribution in zero field of the sublattice with moments along the (vertical) field direction, $(1-10)$. We use the crystal unit cell with eight Yb atoms, and thus there are four Yb moments in each sublattice. We account for this when normalizing the measured magnetic scattering intensity in units of $\mu_B^2 \cdot meV^{-1}/Yb$, as discussed above and shown in figures in the main text and in most of the figures here.

### B. Polarization factor in the $Q_{HH}$ wave vector dependence of magnetic scattering

The energy dependent intensity shown in Fig. S8 E and F was only integrated over $Q_{HH} = [-0.1, 0.1]$, and there is a good reason for that. The total intensity integrated over the full Brillouine zone, $Q_{HH} = [0, 1]$, in 4 T is much less than a half of the total intensity in 0 T integrated over the same $Q-$range. This results from the different polarization factors for the two sublattices, which clearly reveal the longitudinal (with respect to the local Yb Ising axis) character of the measured neutron spectra, as we discuss in this section.

Since neutrons only scatter from magnetic moments that are perpendicular to the wave vector $\boldsymbol{Q}$, the scattering from the two sublattices will have different polarization factors in the (HHL) scattering plane due to their different moment orientations. The moments in the field-dependent (1-10) sublattice directed along magnetic field are vertical and perpendicular to the wave vectors in the scattering plane, which leads to the polarization factor of 1, independent of $\boldsymbol{Q}$. Moments of the perpendicular, (110) sublattice, lie in the scattering plane, and therefore their contribution depends strongly on the alignment of $\boldsymbol{Q}$ with respect to the (110), or, equivalently, (001) direction. Shown in Fig. S9A is the contour map of the inelastic scattering intensity ($E = [0.15, 1.5]$ meV, $Q_{H-H} = [-0.1, 0.1]$) in the (HHL) scattering plane, measured at T = 1.5 K, with the 4.75 T high field data directly subtracted. This is the net inelastic scattering that originates from the sublattice with moments along the (1-10) direction, perpendicular to the (HHL) scattering plane. We observe stripes of intensity that are symmetrically positioned at $Q_{00L} = \pm 1$ and are nearly independent of $Q_{HH}$, consistent with the $Q_{HH}$-independent polarization factor. The inelastic scattering intensity measured at 1.5 K in 4.75 T is rather different, as shown in Fig. S9C. This intensity originates almost entirely from the sublattice with magnetic moments along the (110) direction, which lies in the (HHL) scattering plane. We observe that scattering intensity in 4.75 T decreases continuously with the increasing wave vector $Q_{HH}$.

In Fig. S9E, the inelastic scattering intensity at 1.5 K, integrated within $Q_{00L} = [-0.8, -1.2]$, $Q_{H-H} = [-0.1, 0.1]$ [white dashed region in Fig. S9, A and C] and $E = [0.8, 1]$ meV is plotted for the two sublattices as a function of wave vector $Q_{HH}$. The intensity at B = 4.75 T (red circles) decreases quickly with the increasing $Q_{HH}$, and is well described by the polarization factor for magnetic moment fluctuations along the (110) direction (red solid line),

$$P_{(110)} = 1 - \widetilde{Q}_{110}^2 = 1 - \frac{(a_{110}^* H)^2}{(a_{110}^* H)^2 + (c^* L)^2} = \frac{(c^* L)^2}{(a_{110}^* H)^2 + (c^* L)^2}. \tag{S29}$$

In contrast, for the zero field minus 4.75 T data, where the fluctuating magnetic moments are along (1-10), i.e. perpendicular to the (HHL) plane, the polarization factor is 1. Consistent with this, apart from a small dip around $Q_{HH} = 1$ that we explained above is a consequence of absorption, the integrated intensity of the I(0T) − I(4.75T) data is almost constant along $Q_{HH}$, as indicated by the solid black line.

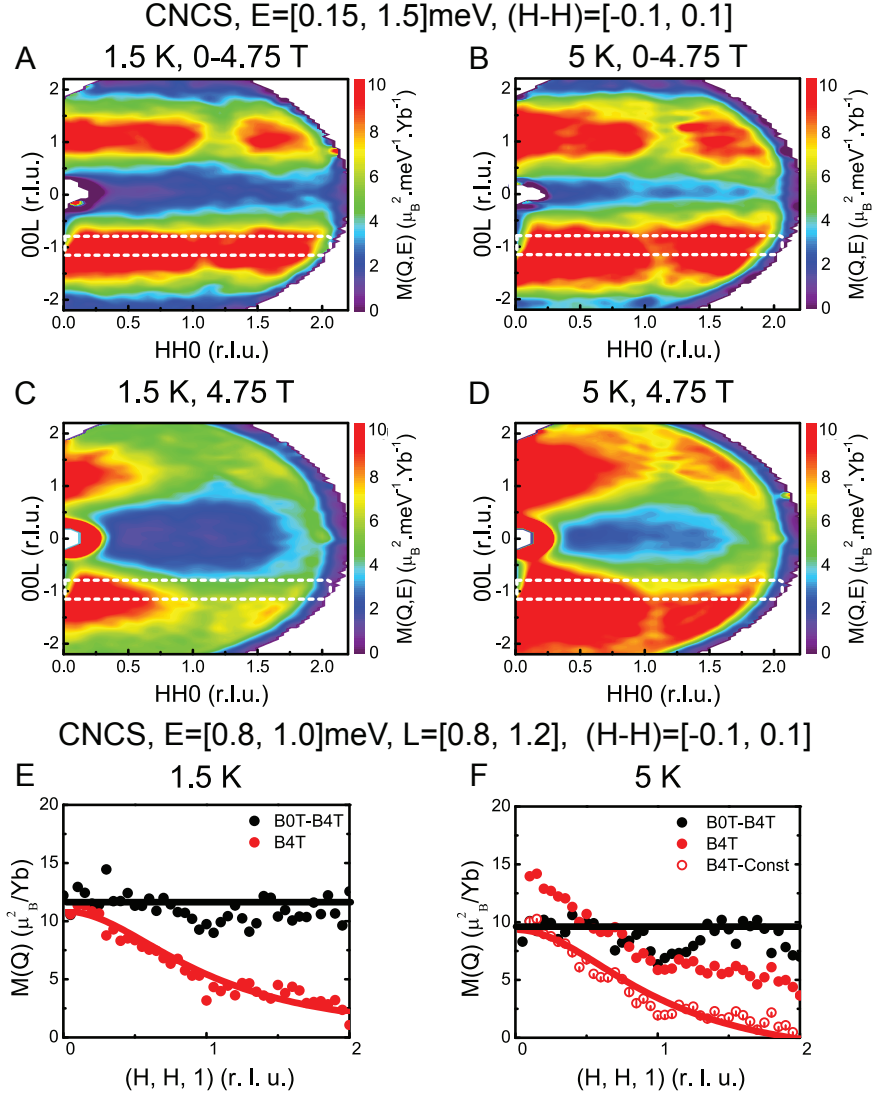

FIG. S9. **Polarization factor and magnetic intensity dependence on** $Q_{HH}$**.** (A), (B) Contour maps of the inelastic scattering intensity ($E = [0.15, 1.5]$ meV, $Q_{H-H} = [-0.1, 0.1]$) in the (HHL) scattering plane measured on CNCS at T = 1.5 K (A) and at T = 5 K (B), with the 4.75 T high field data subtracted. (C), (D) Contour maps of the inelastic scattering intensity ($E = [0.15, 1.5]$ meV, $Q_{H-H} = [-0.1, 0.1]$) in the (HHL) scattering plane measured on CNCS in 4.75 T at T= 1.5 K (C) and at T= 5 K (D). (E), (F) Inelastic scattering intensity integrated within $E = [0.8, 1]$ meV and averaged over $Q_{00L} = [-0.8, -1.2]$, $Q_{H-H} = [-0.1, 0.1]$, as a function of wave vector $Q_{HH}$, at temperature 1.5 K (E) and 5 K (F), for different magnetic fields as indicated in the figure. The white dashed lines in (A), (B), (C) and (D) indicate the region corresponding to the line cuts shown in (E) and (F). The black and red solid lines indicate the different $Q_{HH}$ dependent polarization factors of the two sublattices, as explained in the text. The 4.75 T data in (F) (solid red circles) give the open red circles when a constant is subtracted, as explained in the text.

Data in Fig. S9 B, D and F show the scattering measured in the paramagnetic state at T = 5 K. Both the difference, I(0 T) - I(4.75 T) [Fig. S9B], and the 4.75 T data [Fig. S9D], at 5 K demonstrate very similar $Q_{HH}$ dependencies to those measured at 1.5 K [Fig. S9, A and C]. The nearly $Q_{HH}$ independent stripe of scattering in Fig. S9B and the black solid symbols in Fig. S9F are consistent with the idea that the I(0 T) - I(4.75 T) data represent scattering from the sublattice with moments along the field, (1-10) direction, perpendicular to the (HHL) scattering plane, whose polarization factor is 1, independent of $Q_{HH}$. The $Q_{HH}$ dependence of the 4.75 T data [solid red symbols in Fig. S9F] is similar to that at 1.5 K, up to a $Q_{HH}$-independent constant. This additional constant contribution is not unexpected, and can be explained by the increased importance of thermal fluctuations at 5 K. As a result, B = 4.75 T field appears insufficient to fully polarize the moments of the (1-10) sublattice, and it can still contribute to the inelastic scattering intensity. We believe that this is the origin of the extra $Q_{HH}$ independent scattering at 4.75 T, which we therefore subtract in order to isolate the net contribution from the (110) sublattice [open red circles in Fig. S9F]. As indicated by the red solid line, the resulting $Q_{HH}$ dependence is well described by the same polarization factor of Eq. (S29), as for the 1.5 K data. These observations show that strong Ising anisotropy imposed by the crystal field constrains the moments to fluctuate only along their own Ising axes both in the ordered state, where, as a result, magnetic fluctuations are parallel to the ordered moments, and in the paramagnetic state above the ordering temperature $T_N$ = 2.07 K, where fluctuations remain uniaxial.

That the dependence of magnetic scattering intensity on the wave vector transfer in the $ab-$plane can be explained entirely by the magnetic polarization factor and has no $Q_{HH}$ dependent structure factors, such as would be expected for a dimer, or a ladder, is the most direct evidence that Yb magnetic moments form nearly uncoupled 1D chains, which fluctuate independently.

### C. The temperature dependence of magnetic scattering

In this section, we show how the fluctuating moments change as a function of temperature, below and above the AF ordering temperature $T_N$=2.07 K. Since the experimental temperature range for the DCS experiment is below $T_N$, we will mostly focus on the CNCS data from 1.5 K up to 100 K. Shown in Fig. S10A is the contour map of the spinon dispersion along $Q_{00L}$ with $Q_{HH} = [0, 1]$, $Q_{H-H} = [-0.1, 0.1]$, measured at 1.5 K, with the high field 4.75 T data subtracted. The overall fan shape along the wave vector $Q_{00L}$ is similar that observed at 0.1 K, but the dispersion appears smeared due to thermal fluctuations. The energy dependence of the scattered intensity at 1.5 K summed over $Q_{00L} = [0, \pm 2]$, $Q_{HH} = [0, 1]$ and $Q_{H-H} = [-0.1, 0.1]$ is plotted in Fig. S10B. The spinon continuum centered around 0.5 meV is much broader than the instrumental resolution, which is indicated by the gray area. Integrating the scattered intensity over the first Brillouin zone, we find the fluctuating moment squared of about $(8.4 \pm 1.2)\mu_B^2$/Yb, which is slightly larger than what is found at 0.1 K. The error bars here are dominated by the systematic error, which we estimate based on the difference of the integrated intensity over the wave vector ranges of $Q_{00L} = [0, -2]$ (black circles and line) and $Q_{00L} = [0, 2]$ (red circles and line) [Fig. S10B].

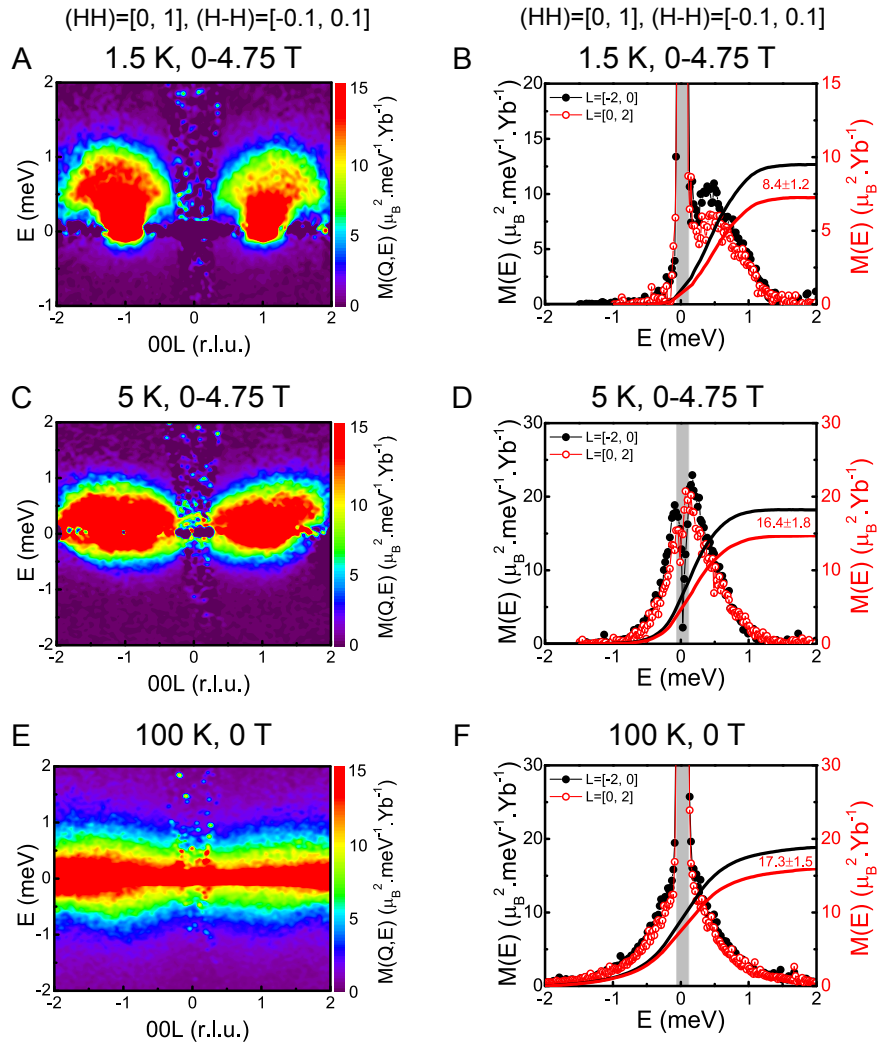

FIG. S10. **Temperature dependence of the inelastic magnetic scattering in Yb$_2$Pt$_2$Pb.** (A) Contour map of the energy dependence of the neutron scattering intensity along $Q_{00L}$ with $Q_{HH} = [0, 1]$, $Q_{H-H} = [-0.1, 0.1]$ at 1.5 K, with the 4.75 T high field data subtracted. (B) The energy dependence of the I(0 T) - I(4.75 T) scattering measured at 1.5 K, summed over $Q_{00L} = [0, \pm 2]$, $Q_{HH} = [0, 1]$ and $Q_{H-H} = [-0.1, 0.1]$. (C) Contour map of the spinon dispersion along $Q_{00L}$ with $Q_{HH} = [0, 1]$, $Q_{H-H} = [-0.1, 0.1]$ at 5 K, with the 4.75 T high field data subtracted. (D) The energy dependence of the I(0 T) - I(4.75 T) scattering measured at 5 K summed over $Q_{00L} = [0, \pm 2]$, $Q_{HH} = [0, 1]$ and $Q_{H-H} = [-0.1, 0.1]$. (E) Contour map of the spinon dispersion along $Q_{00L}$ with $Q_{HH} = [0, 1]$, $Q_{H-H} = [-0.1, 0.1]$, at 100 K in zero field. (F) Energy dependence of the 0 T scattering measured at 100 K, summed over $Q_{00L} = [0, \pm 2]$, $Q_{HH} = [0, 1]$ and $Q_{H-H} = [-0.1, 0.1]$. The gray area in (B), (D), (F) indicates the energy range of elastic diffraction, mainly determined by the instrumental resolution, which is $\approx 0.1$ meV in this experiment.

Once the static order disappears above $T_N$, all the Yb moments are fluctuating. Figure S10C is the contour map of the spinon dispersion measured at 5 K, summed over the same range of wave vectors as at 1.5 K. The overall dispersion along $Q_{00L}$ is now considerably smeared by thermal fluctuations, and the net fluctuating moment obtained by summming the intensity at 5 K over the first Brillouin zone [Fig. S10D] is $(16.4 \pm 1.8)\mu_B^2/$Yb, almost twice the value at 1.5 K. Roughly speaking, about half of the total Yb moments are ordered in the AF state, while the other half remain fluctuating even at the lowest temperatures. This is a very different situation from that found in most rare earth based classical magnets, where spin wave excitations exist only in the presence of magnetic order, and only slightly suppress the magnitude of the ordered moments. The spinon dispersion at 100 K is shown in Fig. S10E. At this temperature, the magnetic field required to fully polarize the sublattice with moments in the field direction is much larger than 4.75 T, which is the largest field available in this experiment. Consequently, we report only zero field measurements. As expected, thermal fluctuations substantially broaden the spectrum for both neutron energy gain and neutron energy loss. The wave vector integrated intensity at 100 K is plotted as a function of energy in Fig. S10F, where the black and red lines are again obtained by integrating the intensity within $Q_{00L} = [0, \pm 2]$. This procedure yields a total fluctuating moment of $(17.3 \pm 1.5)\mu_B^2/$Yb, which is, within our experimental accuracy, identical to the one we found at 5 K. This result is explained by pointing out that little variation in the fluctuating moment is expected as long as the experimental temperature is small compared to the energy splitting, $\Delta_1 \sim 30$meV, of the CEF ground state and the first excited state.

---